\documentclass{aa}  

\usepackage{graphicx}
\usepackage{amsmath}
\usepackage{txfonts}
\usepackage{xcolor}
\usepackage{comment}
\newcommand{\s}{{\rm s}}
\newcommand{\g}{{\rm g}}
\newcommand{\erg}{{\rm erg}}

\newcommand{\cm}{{\rm cm}}

\newcommand{\km}{{\rm km}}
\newcommand{\AU}{{\rm AU}}

\newcommand{\K}{\rm K}
\newcommand{\yr}{\rm yr}

\newcommand{\Msun}{M_{\odot}}
\newcommand{\Mjup}{M_{\rm Jup}}
\newcommand{\Mear}{M_{\oplus}}

\newcommand{\SSP}{SSP}

\begin{document}

   \title{
          %(A) Late planetesimal accretion onto gas giant planets \\
          The origin of the high metallicity of close-in giant exoplanets II
          }

   \subtitle{The nature of the sweet spot for accretion}

   \author{S. Shibata
          \inst{1,2}
          \and
          R. Helled\inst{1}\fnmsep
          \and
          M. Ikoma\inst{3,4,2}
          }

   \institute{Institute for Computational Science (ICS), University of Zurich,
              \email{sho.shibata@uzh.ch}
        \and
            Department of Earth and Planetary Science, Graduate School of Science, The University of Tokyo, 7-3-1 Hongo, Bunkyo-ku, Tokyo 113-0033, Japan
         \and
            Division of Science, National Astronomical Observatory of Japan, 2-21-1 Osawa, Mitaka, Tokyo 181-8588, Japan 
        \and
           Department of Astronomical Science, The Graduate University for Advanced Studies (SOKENDAI), 2-21-1 Osawa, Mitaka, Tokyo 181-8588, Japan 
%             \email{c.ptolemy@hipparch.uheaven.space}
             %\thanks{The university of heaven temporarily does not accept e-mails}
             }

   \date{Received XXX XX, 2021; accepted XXXXX XX, 2021}

% \abstract{}{}{}{}{} 
% 5 {} token are mandatory
 
  \abstract
  %\input{text/abstract}
   {
    The composition of gas giant planets reflects their formation and evolution history. 
    Revealing the origin of the high heavy-element masses in giant exoplanets is a objective of planet  formation theories.
    Planetesimal accretion during the phase of planetary migration could lead to the delivery of heavy elements into gas giant planets. 
    In our previous paper  \citep{Shibata+2020} we used  dynamical simulations and showed that planetesimal accretion during  planetary migration occurs in a rather narrow region of the protoplanetary disk, which we refer as ''the sweet spot for accretion".
    }
   {
    Our understanding of the sweet spot, however, is still limited.  
    The location of the sweet spot within the disk and how it changes as the disk evolves was not investigated in detail.
    The goal of this paper is to reveal the nature of the sweet spot using analytical calculations and investigate the role of the sweet spot in determining the composition of gas giant planets. 
   }
   {
    We analytically derive the required conditions for the sweet spot.
    Then, using the numerical integration of the orbits of planetesimals around a migrating planet, we compare the derived equations with the numerical results. 
   }
   {
    We find that the conditions required for the sweet spot can be expressed by the ratio of the  aerodynamic gas damping timescale of the planetesimal orbits and the planetary migration timescale. 
    %The location of the sweet spot shifts outward with increasing planetary mass because the sweet spot is regulated by the effect of mean motion resonances.
    If the planetary migration timescale depends on the surface density of disk gas inversely, the location of the sweet spot does not change with the disk evolution. 
    We expect that the planets observed inner to the sweet spot include much more heavy elements than the planets outer to that.
    The mass of planetesimals accreted by the protoplanet in the sweet spot depends on the amount of planetesimals that are shepherded by  mean motion resonances. 
    %We also discuss the effects of mutual collisions of planetesimals trapped in the mean motion resonances. % during the migration of a gas giant planet
    Our analysis suggests that tens Earth-mass of planetesimals can be shepherded into the sweet spot without planetesimal collisions.
    However, as more planetesimals are trapped into mean motion resonances, collisional cascade can lead to fragmentation and the production of smaller planetesimals. 
    This could affect the location of the sweet spot and the population of small objects in planetary systems.
    %The mutual collisions of planetesimals in the resonant shepherding might be important for determining the population of small objects in planetary systems.
   }
   {
    We conclude that the composition of gas giant planets depends on whether the planets crossed the sweet spot during their formation. 
    Constraining the metallicity of cold giant planets, that are expected to be outer than the sweet spot, with future observations would reveal key information for understanding the origin of heavy elements in giant planets.
   }

            
   \maketitle
%
%-------------------------------------------------------------------

\section{Introduction}
\label{sec:intro}
The formation of gas giant planets is a complex process which involves many underlying physical processes, such as core formation, gas accretion, and planetary migration.
Recent planetary population synthesis models \citep[e.g.,][]{Ida+2018,Mordasini2018} aim to consider as many processes as possible and construct a  comprehensive planetary formation model which begins at the early stages of planetesimal accretion up to the formation of gas giant planets.
However, there are still many uncertainties related to each of these processes as well as the initial conditions. 
%which are not observable in current planetary systems.
Constraining theories using observed physical parameters is crucial for understanding the formation and evolution  of gas giant planets.
Along with the development of observation and characterisation of exoplanets, many studies aimed to validate theoretically proposed formation models and  constrain the initial parameters that control giant planet formation. 
The planetary chemical composition is a key characteristic that can be used to constrain formation models because the formation history is imprinted in the composition of gas giant planets. %and internal structure 

Recent observations provide information on the chemical composition of gas giant planets. %, in particular, when it comes to giant exoplanets.
Typically it is the planetary metallicity is used for their characterisation and for constraining their formation and evolution histories \citep[e.g.,][]{Helled2021}.  
The composition of gas giant planets was first constrained for the solar-system giant planets Jupiter and Saturn \citep[e.g.][]{Guillot+1999,Saumon+2004}.
From the gravitational moments observed by spacecrafts, the total heavy-element mass in Jupiter $M_{\rm Z,Jup}$ was estimated to be  $\sim25$-$45\Mear$ \citep{Wahl+2017,Debras+2019} and $M_{\rm Z,Sat}\sim16$-$30\Mear$ for Saturn \citep{Saumon+2004,Helled+2013,Movshovitz+2020,Mankovich+2021}. 
Heavy-element enrichment has also been inferred for giant exoplanets  \citep[e.g.,][]{Guillot+2006,Miller+2011,Thorngren+2016}.
According to \citet{Thorngren+2016}, close-in gas giant planets contain large amounts of heavy elements of the order of tens of Earth masses, with  bulk planetary metallicities that are significantly higher than stellar metallicities. 
Surprisingly, some of these planets are estimated to consist of more than $\sim100\Mear$ of heavy elements.
Revealing the origin of this significant heavy-element enrichment in giant planets is a key objective in giant planet formation theory. 

Planetesimal accretion has been proposed to be one of the main sources of heavy-element enrichment of giant planets. 
After the onset of runaway gas accretion, a large amount of planetesimals enters the expanding feeding zone where the planetesimals can be captured by the protoplanet \citep{Zhou+2007,Shiraishi+2008,Shibata+2019,Podolak+2020}.
During the rapid gas accretion phase, up to 30\% of planetesimals inside the feeding zone are captured by the protoplanet \citep{Shibata+2019}.
For a massive protoplanetary disk, this accretion process is sufficient to explain the estimated heavy-element masses in Jupiter and Saturn.
For close-in giant exoplanets,  however, the heavy-element mass that  accreted during rapid gas accretion  is rather low because the total mass planetesimals inside the feeding zone of these planets is significantly lower.

During the planetary migration phase, migrating planets encounter planetesimals in wide regions within the  protoplanetary disk.
Planetesimal accretion during  the migration phase was investigated in the context of accelerating the formation of Jupiter's core \citep[e.g.][]{Tanaka+1999,Alibert+2005}.
It was found that the formation timescale of Jupiter's core can be shortened if planetary migration is included.  
For the origin of large amounts of heavy elements in close-in giant exoplanets, planetesimal accretion by a migrating Jupiter-size planet was also investigated by \citet{Shibata+2020} and \citet{Turrini+2021}.
\citet{Shibata+2020} performed  orbital integration calculations of planetesimals around a migrating planet,  and found that several-tens Earth mass of planetesimals can be accreted by a gas giant planet if the planet migrates tens of AU in a massive protoplanetary disk.
\citet{Turrini+2021} also investigated this process and found that planetesimal accretion during planetary migration is a useful tracer of the formation location of giant planets.
Thus, planetesimal accretion during the planetary migration phase is a key mechanism in the formation history of gas giant planets. 

If the migrating planet is as massive as Jupiter, efficient planetesimal accretion is limited to a narrow ring-like region within the  protoplanetary disk \citep{Shibata+2020}; in this paper, we refer to this region  as the {\it sweet spot for planetesimal accretion} (hereafter, \SSP).
Even in the case of a migrating  Earth-mass planet, the accretion area is limited to an outer disk region (without outer edge) because of the disk gas drag \citep{Tanaka+1999}.  However, such a ring-like region of \SSP\ is specific to the Jupiter-mass planet.
The shape and location of the accretion region 
%are important for determining the accretion efficiency because the total heavy-element mass of captured planetesimals 
during the  planetary migration determine the accretion area and the accretion efficiency. 
The accretion area of planets with  different masses depends on the strength of mean motion resonances accompanied by the migrating planet.
Nevertheless, the detailed nature of \SSP\ is not fully understood.

In this paper, we investigate the nature of \SSP \ in detail and derive analytical equations for determining its location.
In sec.~\ref{sec:sweet_spot}, we analytically investigate the orbital evolution of planetesimals around the migrating planet.
We derive a condition required for planetesimal accretion during the  planetary migration phase.
In sec.~\ref{sec:sweet_spot_numerical}, we compare the derived condition with the numerical simulations.
In sec.~\ref{sec:discussion}, we discuss about the location of \SSP\ in a protoplanetary disk and the heavy-element enrichment of giant planets.
Our conclusions are summarised in Section~\ref{sec:conclusion}.

%--------------------------------------------------------------------
\section{An analytical expression of the accretion sweet spot}
\label{sec:sweet_spot}

\subsection{An overview of planetesimal accretion during planetary migration}\label{sec:overview}
%%%%%%%%%%%%%%%%%%%%%%%%%%%%%%%%%%%%%%%%%%%%%%%%%%%%%%%%%%%%%
\begin{figure}
  \begin{center}
    \includegraphics[width=90mm]{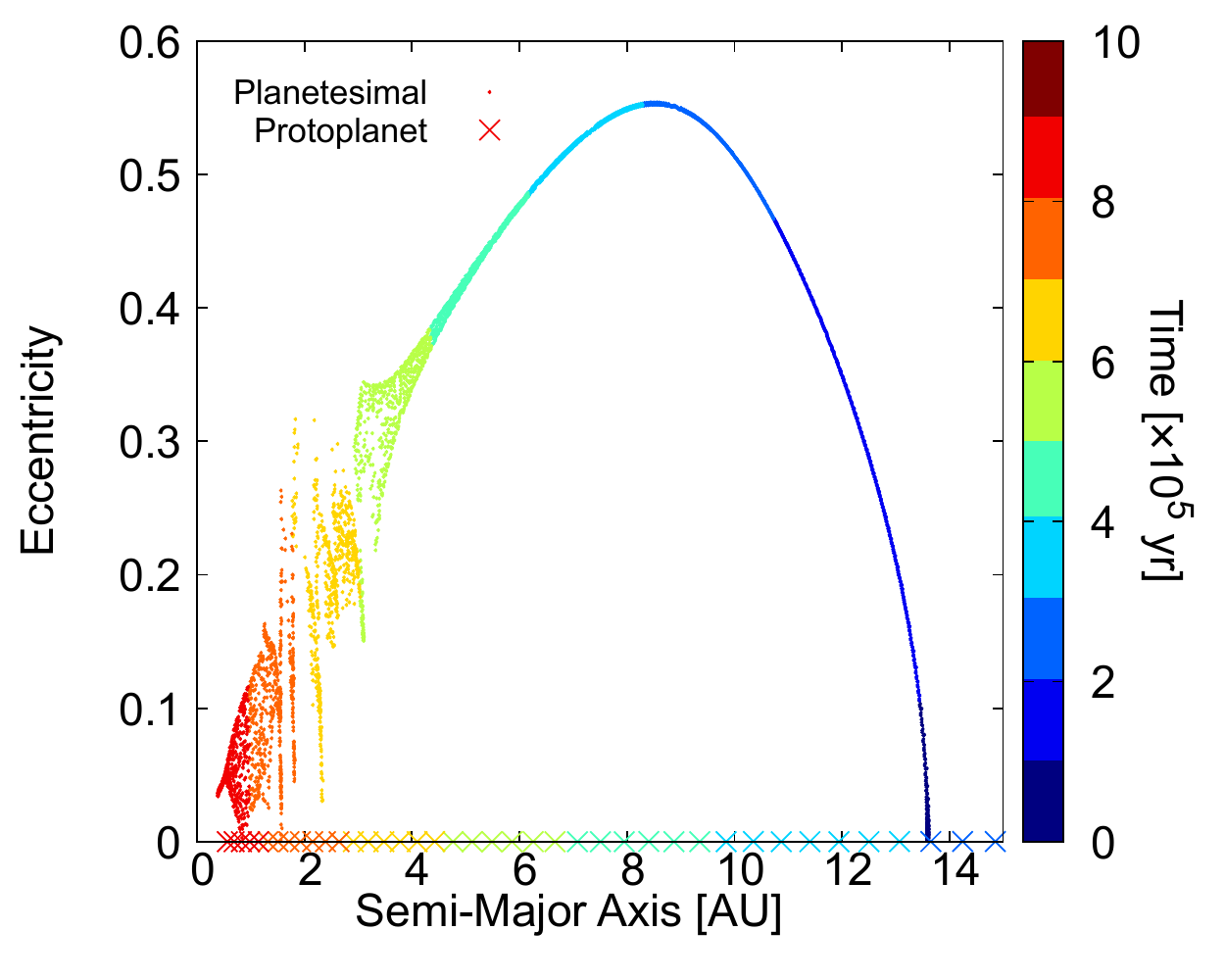}
    \caption{
    Orbital evolution of a planetesimal initially located at $13.8\AU$ dictated via the gravitational interaction with a migrating Jupiter-mass protoplanet. 
    The dots and crosses show the locations of the planetesimal and the protoplanet, respectively, and are colour-coded according to time.
    In this simulation the surface density of disk gas is 
    set to that of the minimum-mass solar nebula and aerodynamic gas drag is included. 
    Here the protoplanet migrates from $15.0\AU$ with migration rate of $10^5$ AU/yr.
    For the details of the numerical simulation, see sec.~\ref{sec:sweet_spot_numerical}.
    }
    \label{fig:axi_ecc}
  \end{center}
\end{figure}
%%%%%%%%%%%%%%%%%%%%%%%%%%%%%%%%%%%%%%%%%%%%%%%%%%%%%%%%%%%%%
Mean motion resonances play an important role in the accretion of planetesimals onto a migrating protoplanet.
Here, we summarise the orbital evolution of planetesimals dictated via their gravitational interaction with a migrating protoplanet, which as found by the numerical simulations of \citet{Shibata+2020}.
Figure~\ref{fig:axi_ecc} shows the orbital evolution of a planetesimal initially located at $13.8\AU$ around a migrating Jupiter-mass protoplanet. 
The orbital evolution of the planetesimal is divided into three characteristic phases as discussed below.

First phase ($t=0~\yr$ -- $4\times10^5~\yr$; bluish symbols): 
The planetesimal encounters the migrating protoplanet and is trapped in the $2:1$ mean motion resonance; this phenomenon is referred to as the \textit{resonant trapping}.
The planetesimal trapped in the mean motion resonance is transported inward together with the migrating protoplanet and its eccentricity is highly enhanced.
This phenomenon is known as the {\it resonant shepherding} \citep[][]{Batygin+2015}.

Second phase ($t=4\times10^5~\yr$ -- $7\times10^5~\yr$; green-yellow symbols): 
As time progresses the resonantly-trapped planetesimal starts to escape from the mean motion resonance, and its eccentricity begins to decrease.
This is because of the stronger aerodynamic drag in the inner regions.
The break of resonant trapping shown in Fig.~\ref{fig:axi_ecc} is caused by overstable liblation \citep{Goldreich+2014}.
The eccentricity of the trapped planetesimal is excited through gravitational scattering by the migrating protoplanet but is also damped by the disk gas drag. 
This overstable equilibrium condition makes the planetesimal orbit unstable and breaks the resonant trapping.

Third phase ($t=7\times10^5~\yr$ --; reddish symbols): 
In the farther inner region, the disk gas is dense enough that the resonantly-trapped planetesimal is damped faster than the planetary migration and, therefore, the planetesimal is eliminated from the feeding zone.
We refer to this phenomenon as the {\it aerodynamic shepherding}, which was found by \citet{Tanaka+1999} in the context of terrestrial planet formation. 

According to the numerical simulations preformed in \citet{Shibata+2020}, planetesimal accretion occurs during the second phase, and the region in which for efficient planetesimal accretion occurs is referred to as \SSP. 
During the first and third phases, planetesimal accretion is negligible.  
The existence of these three different phases (or regions) is associated with the change in strength of the aerodynamic gas drag which increases as the protoplanet migrates inward.

In the following sections, we derive the conditions for the \SSP\ analytically. 
In Sec.~\ref{sec:condition_for_resonant_trapping}, we show that planetesimals are generally trapped in the mean motion resonances with an approaching protoplanet in standard protoplanetary disks. 
In Sec.~\ref{sec:orbital_evolution_in_resonant_trapping}, we show that the eccentricity of the resonantly-trapped planetesimal reaches an equilibrium value, $e_{\rm eq}$, that is determined by the ratio of  the aerodynamic damping timescale (or the friction timescale) for the planetesimal ($\tau_{\rm aero, 0}$; Eq.~[\ref{eq:aerodynamic_damping_timescale}]) and the tidal damping timescale (or the migration timescale) of the protoplanet ($\tau_{\mathrm{tide}, a}$). 
A necessary condition for a planetesimal to be captured is that it enters the feeding zone of the protoplanet. 
As for the resonantly trapped planetesimal, $e_{\rm eq}$ must exceed the eccentricity at the feeding zone boundary ($e_{\rm cross}$; Eq.~[\ref{eq:ecc_cross_resonance_feeding_zone}]), which is derived in Sec.~\ref{sec:Condition_for_Aerodynamic_Shepherding}. 
Even if such a condition is satisfied, however, for the trapped planetesimal to escape the resonance completely, the perturbed planetesimal's orbit must oscillate more widely relative to the resonant width (i.e., the overstable libration) and such an oscillation must be amplified more rapidly than the protoplanet's migration.
The conditions are derived in Sec.~\ref{sec:condition_for_resonant_breaking}. 
Finally, by putting everything above together, we infer the condition for the \SSP  \, in Sec.~\ref{sec:condition_for_sweet_spot}.

\subsection{Before resonant trapping}\label{sec:condition_for_resonant_trapping}
When the radial distance between two objects decreases, which is referred to as convergent orbital evolution, the objects are eventually locked in a resonant state.
Such resonant trapping has been discussed in the contexts of the formation of satellite pairs \citep[e.g.,][]{Goldreich1965,Dermott+1988,Malhotra1993b}, the transport of small bodies \citep[e.g.,][]{Yu+2001,Batygin+2015}, and the formation of exoplanet pairs \citep[e.g.,][and references therein]{Fabrycky+2014,Goldreich+2014,Batygin+2015b}. 
%; see also references in \citet{Batygin+2015b}.
However, not all resonant encounters result in resonant trapping. 
The resonant trapping requires the following conditions for a planetesimal-protoplanet pair \cite[e.g.][]{Malhotra1993}:
\begin{itemize}
    \item[(i)] The two orbits converge with each other;
    \item[(ii)] The eccentricity of the planetesimal, $e$, before it is trapped in the resonance is smaller than the critical value $e_{\rm crit}$ (see Eq.~(\ref{eq:critical_eccentricity}));
    \item[(iii)] The timescale for the planetesimal to cross the resonant width $\tau_{\rm cross}$ is longer than the libration timescale $\tau_{\rm lib}$ (see Eqs.~(\ref{eq:crossing_timescale}) and (\ref{eq:1st_order_libration_time_in})).
\end{itemize}
Here, we consider a trap of a planetesimal in the inner first-order mean motion resonances ($j:j-1$ resonance) with a protoplanet migrating inward and assume the mass of the planetesimal is negligibly small relative to %the mass 
that of the protoplanet.

The first condition can be obtained by comparing the timescales of change in the semi-major axis of the planetesimal and the protoplanet.
In a protoplanetary gaseous disk, the orbits of objects shrink by the aerodynamic gas drag and gravitational tidal drag from the disk gas.
The former drag is dominant for planetesimal-size objects ($\lesssim10^{24}\g$) and the latter drag is for planet-size objects ($\gtrsim10^{24}\g$) \citep{Zhou+2007}.
Using the aerodynamic damping timescale for semi-major axis $\tau_{\mathrm{aero},a}$ for the planetesimal and the tidal damping timescale for semi-major axis $\tau_{{\rm tide},a}$ for the protoplanet, we can write the first condition as: 
\begin{align}
    \tau_{{\rm aero},a} > \tau_{{\rm tide},a}. \label{eq:condition_converge_orbits}
\end{align}

As for the second condition, the critical eccentricity for the $j:j-1$ resonances is given by \citep[e.g.][]{Murray+1999}:
\begin{align}
    e_{\rm crit} = \sqrt{6} \left[ \frac{3}{|f_{\rm d}|} \left(j-1\right)^{4/3} j^{2/3} \frac{M_{\rm s}}{M_{\rm p}} \right]^{-1/3}, \label{eq:critical_eccentricity}
\end{align}
where $M_{\rm p}$ and $M_{\rm s}$ are the masses of the planet and central star, respectively, and $f_{\rm d}$ is the interaction coefficient whose values are summarised in Table~\ref{tab:interaction_coefficient}.

When $\tau_{{\rm aero},a} \gg \tau_{{\rm tide},a}$, the timescale for the planetesimal to cross the resonant width $\tau_{\rm cross}$ is given by: 
\begin{align}
    \tau_{\rm cross} = \left| \frac{\Delta a_{\rm res}}{\dot{a}_{\rm c}} \right| = \left| \frac{\Delta a_{\rm res}}{a_{\rm c}} \right| \tau_{{\rm tide},a}, \label{eq:crossing_timescale}
\end{align}
where $a_{\rm c}$ is the semi-major axis of the resonance centre, which is related to the protoplanet's semi-major axis $a_{\rm p}$ as: 
\begin{align}
    a_{\rm c} &= \left(\frac{j-1}{j}\right)^{2/3} a_{\rm p}, \label{eq:resonant_center}
\end{align}
$\dot{a}_{\rm c}$ is the temporal change of $a_{\rm c}$, and $\Delta a_{\rm res}$ is the width of the resonance given by \citep[e.g.][]{Murray+1999}: 
\begin{align}
    \frac{\Delta a_{\rm res}}{a_{\rm c}} &= 2 \left\{ \frac{16}{3} \frac{M_{\rm p}}{M_{\rm s}} \left( \frac{j-1}{j} \right)^{2/3} e \left|f_{\rm d} \right| \right\}^{1/2} \nonumber \\
    & \qquad \left\{ 1+\frac{1}{27 (j-1)^2 e^3} \frac{M_{\rm p}}{M_{\rm s}} \left( \frac{j-1}{j} \right)^{2/3} \left| f_{\rm d} \right| \right\}^{1/2}. \label{eq:1st_order_resonant_width} 
\end{align}
%Note that the resonance centre migrates with the same rate of the protoplanet during the planetary migration phase.
The width of the resonance depends on the eccentricity and takes the minimum value at: 
\begin{align}
    e = \frac{2^{1/3}}{3} \left( \left|j \right|^{-2/3} \left|j-1\right|^{-4/3} \left| f_{\rm d} \right| \frac{M_{\rm p}}{M_{\rm s}} \right)^{1/3}. \label{eq:ecc_at_minimum_resonant_width}
\end{align}
By substituting Eq.~(\ref{eq:ecc_at_minimum_resonant_width}) into Eq.~(\ref{eq:1st_order_resonant_width}), we obtain the minimum width of the resonance as:
\begin{align}
    \left. \frac{\Delta a_{\rm res}}{a_{\rm c}} \right|_{\rm min} = \frac{2^{8/3}}{3^{1/2}} j^{-4/9} (j-1)^{1/9} {|f_{\rm d}|}^{2/3} \left( \frac{M_{\rm p}}{M_{\rm s}} \right)^{2/3}. \label{eq:minimum_resonant_width}
\end{align} 
The libration timescale at internal resonances are given by \citep[e.g.][]{Murray+1999,Batygin+2015}:
\begin{align}
    \tau_{\rm lib}  = \frac{2 \pi}{n} &\left| \left( \frac{j-1}{j} \right)^{4/3} \left( \frac{M_{\rm p}}{M_{\rm s}} \right)^2 \frac{{f_{\rm d}}^2 }{e^2} \right. \nonumber \\
                    & \left.  - 3 (j-1)^2 \left(\frac{j-1}{j} \right)^{2/3} \frac{M_{\rm p}}{M_{\rm s}} f_{\rm d} e \right|^{-1/2}, \label{eq:1st_order_libration_time_in}
\end{align}
where $n$ is the planetesimals' mean motion. 
Substituting Eq.~(\ref{eq:ecc_at_minimum_resonant_width}) into Eq.~(\ref{eq:1st_order_libration_time_in}), we obtain the corresponding libration timescale as: 
\begin{align}
    \tau_{\rm lib} = 0.48 j^{4/9} (j-1)^{-10/9} \left( \frac{M_{\rm p}}{M_{\rm s}} \right)^{-2/3} |f_{\rm d}|^{-2/3} T_{\rm K} \label{eq:corresponding_libration_timescale}.
\end{align}
where $T_{\rm K}$ ($= 2 \pi / n$) is the Kepler period at the resonance centre.
Using Eqs.~(\ref{eq:crossing_timescale}), (\ref{eq:minimum_resonant_width}) and (\ref{eq:corresponding_libration_timescale}), we obtain the third (sufficient) condition as
\begin{align}
    \frac{\tau_{\rm cross}}{\tau_{\rm lib}} = 7.7 j^{1/9} \left(j-1\right)^{2/9} |f_{\rm d}|^{4/3} \left(\frac{M_{\rm p}}{M_{\rm s}}\right)^{4/3} \frac{\tau_{{\rm tide},a}}{T_{\rm K,p}} \gtrsim 1, \label{eq:condition_for_resonant_trapping_mig}
\end{align}
where $T_{\rm K,p}$ is the Kepler period of the migrating protoplanet; namely, $T_{\rm K, p}$ = $T_{\rm K} \, j/(j-1)$.

\begin{figure}
  \begin{center}
    \includegraphics[width=90mm]{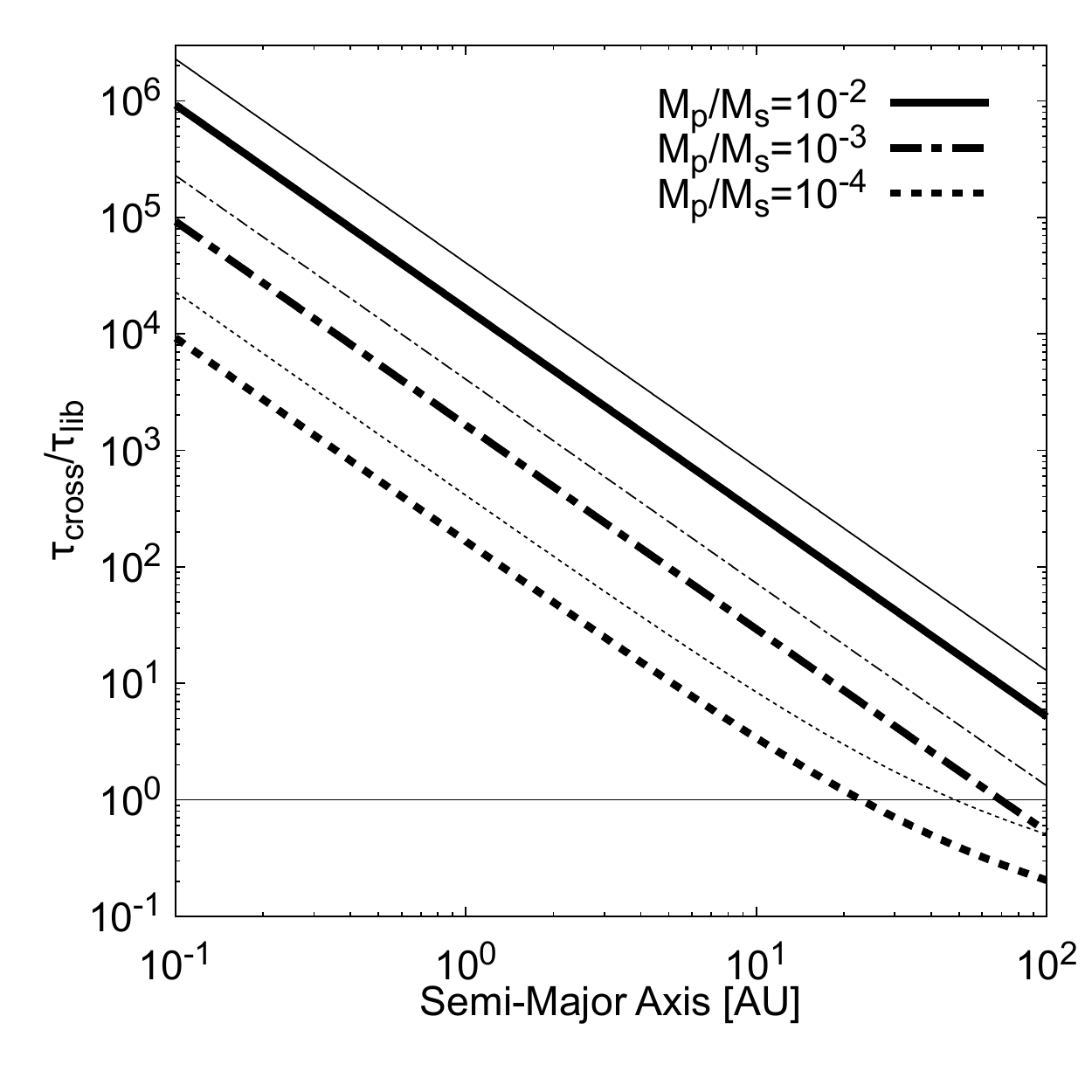}
    \caption{
    The ratio between the crossing timescale $\tau_{\rm cross}$ (Eq.~\ref{eq:crossing_timescale}) and the libration timescale $\tau_{\rm lib}$ (Eq.~\ref{eq:corresponding_libration_timescale}) as a function of the planetesimal's semi-major axis. 
    Here, we consider the type II regime obtained in \citet{Kanagawa+2018} for the migration timescale $\tau_{{\rm tide},a}$.
    Solid, dash-dotted, and dotted lines show the cases of the planet-to-star mass ratio $M_{\rm p}/M_{\rm s}=10^{-2}$, $10^{-3}$ and $10^{-4}$, respectively. 
    We consider the inner first-order mean motion resonances (or $j$:$j-1$ resonances); thick lines are for $j=2$ and thin lines are for $j=3$.
    To calculate the migration timescale, we used a disk model same as that in \citet{Ida+2018}. 
    The disk accretion rate, disk's viscosity parameter, and disk aspect ratio are set to $10^{-8} M_{\odot}/\yr$, $10^{-3}$ and $0.03 r^{1/4}$, respectively.
    }
    \label{fig:Timescales_for_resonant_trapping}
  \end{center}
\end{figure}

Far from the protoplanet and outside mean motion resonances, the planetesimal's eccentricity is determined by the balance between the viscous stirring from the planetesimal swarm and the aerodynamic gas drag. 
The mean value of r.m.s. eccentricities of the planetesimal swarm is on the order of $\lesssim10^{-2}$ \citep{Ohtsuki+2002}. %[e.g.][]
For km-sized planetesimals with $e=10^{-2}$, the radial drift timescale $\tau_{{\rm damp},a}$ is longer than the planetary migration timescale in the type II regime ($\tau_{{\rm tide},a}\lesssim10^7~\yr$) over a wide region of a protoplanetary disk \citep{Adachi+1976}. 
Also, the critical eccentricity is $e_{\rm crit}\sim0.15$ for $j=2$ and $M_{\rm p}/M_{\rm s}=10^{-3}$.
Thus, the first and second conditions are expected to be achieved in protoplanetary disks.
Figure~\ref{fig:Timescales_for_resonant_trapping} shows the timescale ratio $\tau_{\rm cross}/\tau_{\rm lib}$ as a function of  semi-major axis, calculated from Eq.~(\ref{eq:condition_for_resonant_trapping_mig}). %Eqs.~(\ref{eq:crossing_timescale}) and (\ref{eq:corresponding_libration_timescale}).
Here, we show the cases of $2:1$ (thick lines) and $3:2$ (thin lines) resonances, and consider the type II regime for the migration timescale $\tau_{{\rm tide},a}$ obtained in \citet{Kanagawa+2018}. 
To calculate $\tau_{{\rm tide},a}$, we adopt the self-similar solution of \citet{Lynden-Bell+1974} for the disk structure with a disk accretion rate of $10^{-8} M_{\odot}/\yr$ and viscosity of $10^{-3}$.
The disk aspect ratio is set to $0.03 r^{1/4}$. 
The timescales ratio $\tau_{\rm cross}/\tau_{\rm lib}$ is larger than unity except for the cases of $M_{\rm p}/M_{\rm s}= 10^{-4}$ in outer disk $a \gg 10 \AU$.
Thus, we conclude that the resonant trapping of planetesimals usually occurs during the migration of a giant protoplanet.

%%%%%%%%%%%%%%%%%%%%%%%%%%%%%%%%%%%%%%%%%%%%%%%%%%%%%%%%%%%%%
\begin{table}[t]
    \centering
    \begin{tabular}{c|c|c} \hline
        $p+q:p$ & $f_{\rm s}$ & $f_{\rm d}$  \\ \hline
        2:1     & $0.387627$ & $-1.19049$ \\
        3:2     & $1.15279$  & $-2.02521$ \\
        4:3     & $2.27923$  & $-2.84042$ \\
        5:4     & $3.76541$  & $-3.64962$ \\ \hline
    \end{tabular}
    \caption{Values of the secular term $f_{\rm s}$ and the direct term $f_{\rm d}$ for the first order mean motion resonances.}
    \label{tab:interaction_coefficient}
\end{table}
%%%%%%%%%%%%%%%%%%%%%%%%%%%%%%%%%%%%%%%%%%%%%%%%%%%%%%%%%%%%%

%%%%%%%%%%%%%%%%%%%%%%%%%%%%%%%%%%%%%%%%%%%%%%%%%%%%%%%%%%%%%
%%%%%%%%%%%%%%%%%%%%%%%%%%%%%%%%%%%%%%%%%%%%%%%%%%%%%%%%%%%%%

\subsection{Orbital evolution under resonant trapping}\label{sec:orbital_evolution_in_resonant_trapping}
Trapped planetesimals suffer from gravitational perturbation from the migrating protoplanet and aerodynamic gas drag from the disk gas. 
The Lagrange's equations under the aerodynamic gas drag are given by \citep[e.g.,][]{Murray+1999}:
\begin{align}
    \dot{a} &= 2 (j-1) C_{\rm r} a e   \sin \varphi -\frac{a}{\tau_{{\rm aero},a}}, \label{eq:Lagrange_axi_wGas} \\
    \dot{e} &=       - C_{\rm r}       \sin \varphi -\frac{e}{\tau_{{\rm aero},e}}, \label{eq:Lagrange_ecc_wGas}
\end{align}
where $\varphi$ is the resonant angle and
\begin{align}
    C_{\rm r} =  \frac{M_{\rm p}}{M_{\rm s}} n \alpha f_{\rm d};  \label{eq:coefficient_resonance}
\end{align} 
$\tau_{{\rm aero},a}$ and $\tau_{{\rm aero},e}$ are the aerodynamic damping timescales for semi-major axis and eccentricity, which are given by \citep{Adachi+1976}:
\begin{align}
    \tau_{{\rm aero},a} &= \frac{\tau_{\rm aero,0}}{2} \left\{ (0.97 e +0.64 i +   |\eta_{\rm gas}|)\eta_{\rm gas}  \right\}^{-1}, \label{eq:aerodynamic_damping_axi} \\
    \tau_{{\rm aero},e} &=       \tau_{\rm aero,0}     \left(   0.77 e +0.64 i +1.5|\eta_{\rm gas}| \right)^{-1}, \label{eq:aerodynamic_damping_ecc}
\end{align}
where 
\begin{align}
    \tau_{\rm aero,0} = \frac{2 m_{\rm pl}}{C_{\rm d} \pi {R_{\rm pl}}^2 \rho_{\rm gas} v_{\rm K}}. \label{eq:aerodynamic_damping_timescale}
\end{align}
Here $C_{\rm d}$ is the non-dimensional drag coefficient, 
$m_{\rm pl}$ is the planetesimal's mass, 
$R_{\rm pl}$ is the planetesimal's radius, 
$\rho_{\rm gas}$ is the gas density, and 
$v_{\rm K}$ is the Kepler velocity.
The planetesimal's mass $m_{\rm pl}$ is calculated as $4\pi\rho_{\rm pl}{R_{\rm pl}}^3/3$, where $\rho_{\rm pl}$ is the planetesimal's mean density.% and given as $\rho_{\rm pl}=2 \g/\cm^3$.
$\eta_{\rm gas}$ is a parameter that defines the sub-Keplerian velocity of disk gas as: 
\begin{align}\label{eq:Disk_Gas_Velocity_sub-Kepler}
	v_{\rm gas} = v_{\rm K} \left( 1- \eta_{\rm gas} \right).
\end{align}
Note that we have neglected the high-order terms in eccentricity in Eqs.~(\ref{eq:Lagrange_axi_wGas}), (\ref{eq:Lagrange_ecc_wGas}), (\ref{eq:aerodynamic_damping_axi}) and (\ref{eq:aerodynamic_damping_ecc}).
Although the eccentricities of the trapped planetesimals are easily excited to $e^2 \sim 0.1$, the location of the sweet spot analytically derived with these equations is found to be  consistent with the numerical results, as shown in section~\ref{sec:sweet_spot_numerical}.

During the resonant trapping, the period ratio between the protoplanet and trapped planetesimal remains constant, by definition, and
%to be
%\begin{align}
%    \frac{j-1}{j} &= \frac{n_{\rm p}}{n} = \left( \frac{a}{a_{\rm p}} \right)^{3/2} = \text{const.}, \label{eq:Period_Ratio_in_Resonances}
%\end{align}
%where $n_{\rm p}$ is the planet's mean motion.
%Taking a time derivative of eq.~(\ref{eq:Period_Ratio_in_Resonances}), we find
\begin{align}
    \frac{a}{\dot{a}} = \frac{a_{\rm p}}{\dot{a_{\rm p}}} = -\tau_{{\rm tide},a}. \label{eq:Timescles_in_resonant_trapping}
\end{align}
%Using eqs.~(\ref{eq:Lagrange_axi_wGas}), (\ref{eq:Lagrange_ecc_wGas}), (\ref{eq:Timescles_in_resonant_trapping}) and $\tau_{{\rm damp},a}~\gg~\tau_{{\rm tide},a}$, we obtain
%\begin{align}
%    \dot{e}     &=  \frac{1}{2(j-1)e \tau_{{\rm tide},a}} -\frac{e}{\tau_{{\rm damp},e}}, \label{eq:Solution_Lagrange_ecc} \\
%    \sin \varphi   &= -\frac{1}{2(j-1) C_{\rm r} e \tau_{{\rm tide},a}}. \label{eq:Solution_Lagrange_phi}
%\end{align}
When $\tau_{{\rm aero},a} \gg \tau_{{\rm tide},a}$, using Eq.~(\ref{eq:Timescles_in_resonant_trapping}), we can re-write  Eq.~(\ref{eq:Lagrange_axi_wGas}) as: 
\begin{align}
    \sin \varphi   &= -\frac{1}{2(j-1) C_{\rm r} e \tau_{{\rm tide},a}}. \label{eq:Solution_Lagrange_phi}
\end{align}
Also, using Eq.~(\ref{eq:Solution_Lagrange_phi}) in Eq.~(\ref{eq:Lagrange_ecc_wGas}), we obtain: 
\begin{align}
    \dot{e}     &=  \frac{1}{2(j-1)e \tau_{{\rm tide},a}} -\frac{e}{\tau_{{\rm aero},e}}. \label{eq:Solution_Lagrange_ecc} 
\end{align}
Just after trapped into the resonances, $e$ is small and $\dot{e}$ is positive.
Equation~(\ref{eq:Solution_Lagrange_ecc}) indicates that $e$ increases on a timescale $\sim~e^2~\tau_{{\rm tide},a}~(\ll~\tau_{{\rm tide},a})$ and reaches an equilibrium value.
Solving Eqs.~(\ref{eq:Solution_Lagrange_phi}) and (\ref{eq:Solution_Lagrange_ecc}) with $\dot{e}=0$, we obtain the equilibrium eccentricity and resonant argument given by: 
\begin{align}
    e_{\rm eq}          &= \left\{ \frac{1}{2(j-1)} \frac{\tau_{{\rm aero},e}}{\tau_{{\rm tide},a}} \right\}^{1/2}, \label{eq:equilibrium_eccentricity} \\
    \sin \varphi_{\rm eq}  &= - \frac{e_{\rm eq}}{C_{\rm r} \tau_{{\rm aero},e}}. \label{eq:equilibrium_resonant_argument}
\end{align}
Unlike the eccentricity, the planetesimals' inclinations are not excited by the migrating protoplanet because the planetesimals are far from the protoplanet.
When $e \gg i$ and $e \gg \eta_{\rm gas}$, $\tau_{\mathrm{aero},e} = \tau_{\rm aero,0} (0.77 e)^{-1}$ from Eq.~(\ref{eq:aerodynamic_damping_ecc}) and, thus, $e_{\rm eq}$ and $\sin \varphi_{\rm eq}$ are written approximately as: 
\begin{align}
    e_{\rm eq}          &\sim \left\{ \frac{1}{1.54 (j-1)} \frac{\tau_{\rm aero,0}}{\tau_{{\rm tide},a}} \right\}^{1/3}, \label{eq:equilibrium_eccentricity_app} \\
    \sin \varphi_{\rm eq}  &\sim -\frac{0.77{e_{\rm eq}}^2}{C_{\rm r} \tau_{\rm aero,0}}. \label{eq:equilibrium_resonant_argument_app}
\end{align}
Note that if the high-order terms of the  eccentricity are considered, the equilibrium eccentricity is smaller than that given by Eq.~(\ref{eq:equilibrium_eccentricity_app}). % for $e\gtrsim0.3$.

\subsection{Aerodynamic shepherding}\label{sec:Condition_for_Aerodynamic_Shepherding}
\begin{figure}
  \begin{center}
    \includegraphics[width=80mm]{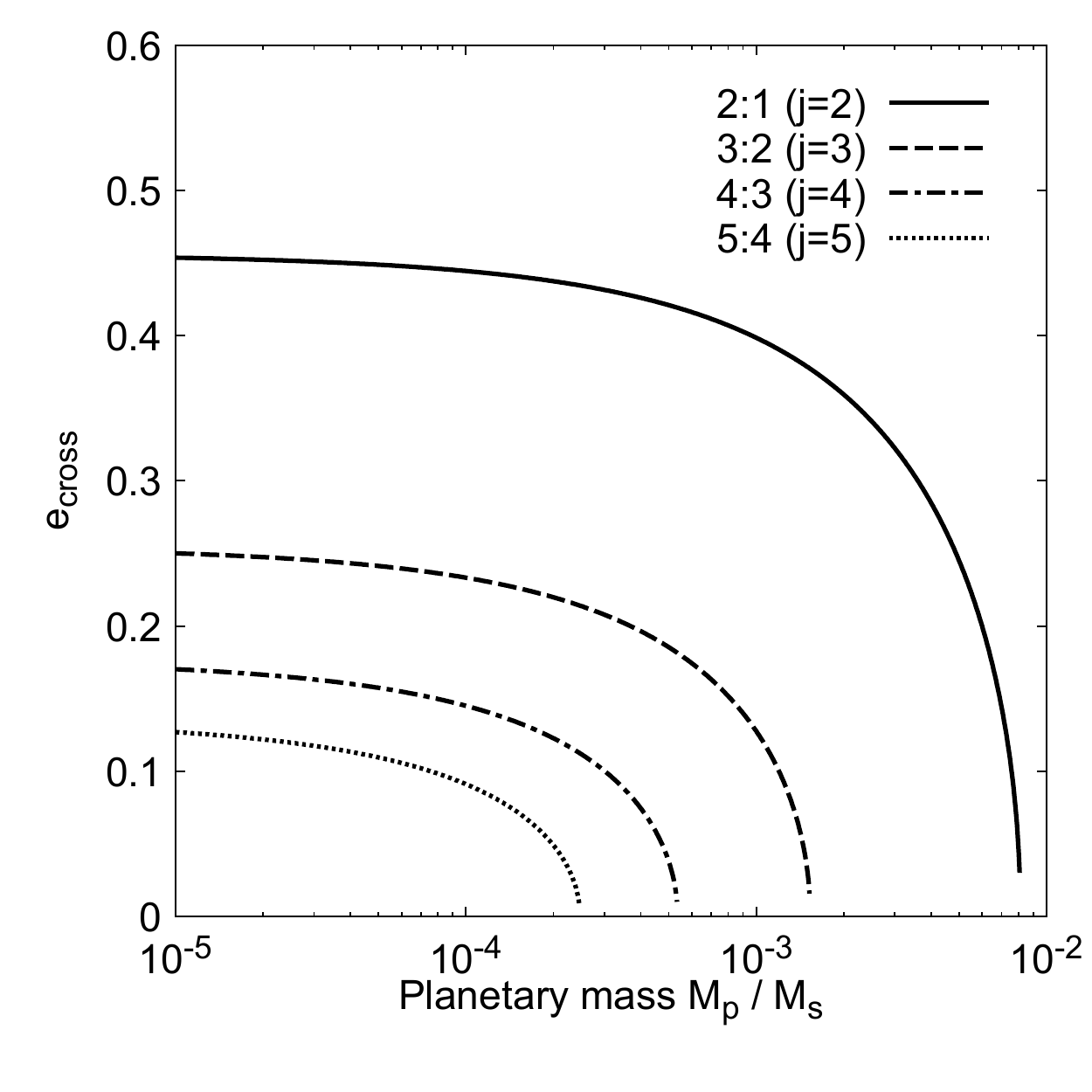}
    \caption{
    The dependence of the crossing point eccentricity $e_{\rm cross}$ (see Eq.~\ref{eq:ecc_cross_resonance_feeding_zone}) on the planet-to-star mass ratio $M_{\rm p}/M_{\rm s}$ for different four cases of $j=2$ (solid), $j=3$ (dashed), $j=4$ (dash-dotted) and $j=5$ (dotted), respectively.
    }
    \label{fig:Mp_EccCross}
  \end{center}
\end{figure}

As shown in \citet{Shibata+2020}, the mean motion resonances play a role like narrow flow channels for planetesimals to enter the planet's feeding zone; the channels are called the {\it accretion bands}. 
Mathematically speaking, planetesimals with negative values of the Jacobi energy, $E_{\rm Jacobi}$, can enter the feeding zone, once their eccentricities are sufficiently enhanced so that their Jacobi energy becomes positive.
The Jacobi energy is given by \citep{Hayashi+1977}: 
\begin{align}\label{eq:Ejacobi}
    E_{\rm Jacobi} = \frac{\mathcal{G} M_{\rm s}}{a_{\rm p}} \left\{ -\frac{a_{\rm p}}{2 a} - \sqrt{\frac{a}{a_{\rm p}} \left(1-e^2\right)} \cos {i} + \frac{3}{2} + \frac{9}{2} h^2 +O(h^3) \right\},
\end{align}
where $h$ is the reduced Hill radius defined as 
\begin{align}\label{eq:reduced_Hill_radius}
    h = \left( \frac{M_{\rm p}}{3 M_{\rm s}} \right)^{1/3}.
\end{align}
Substituting $E_{\rm jacobi}=0$ and $a/a_{\rm p} = \left\{(j-1)/j\right\}^{2/3}$ into Eq.~(\ref{eq:Ejacobi}) and assuming $\cos^2 i \sim 1$, one obtains the eccentricity of the accretion band, $e_{\rm cross}$, as: 
\begin{align}
    e_{\rm cross} = \left[ 1- \left( \frac{j}{j-1} \right)^{2/3} \left\{ \frac{3}{2} +\frac{9}{2} h^2 -\frac{1}{2} \left( \frac{j}{j-1} \right)^{2/3} \right\}^2 \right]^{1/2}. \label{eq:ecc_cross_resonance_feeding_zone}
\end{align}
The condition required for entering the feeding zone is that the equilibrium eccentricity $e_{\rm eq}$ (see Eq.~\ref{eq:equilibrium_eccentricity_app}) is larger than $e_{\rm cross}$, which is given by: 
\begin{align}
    1.54 (j-1) {e_{\rm cross}}^3 &< \frac{\tau_{\rm aero,0}}{\tau_{{\rm tide},a}}. \label{eq:condition_for_aerodynamic_shepherding}
\end{align}
Figure~\ref{fig:Mp_EccCross} shows $e_{\rm cross}$ as a function of $M_{\rm p}/M_{\rm s}$ for different values of $j$ from Eq.~(\ref{eq:ecc_cross_resonance_feeding_zone}).
As indicated in this figure, $e_{\rm cross}$ decreases with increasing planetary mass. 
This is because as the protoplanet mass increases, the feeding zone expands and, consequently, the crossing point (i.e., the accretion band) itself is located inside the feeding zone, once the planetary mass exceeds a certain value.
For example, while there are three crossing points $(j=2,3,4)$ when $M_{\rm p}/M_{\rm s}=10^{-3.5}$, there is only one crossing point $(j=2)$ when $M_{\rm p}/M_{\rm s}=10^{-2.5}$ and no points when $M_{\rm p}/M_{\rm s}=10^{-2}$.

%%%%%%%%%%%%%%%%%%%%%%%%%%%%%%%%%%%%%%%%%%%%%%%%%%%%%%%%%%%%%
%%%%%%%%%%%%%%%%%%%%%%%%%%%%%%%%%%%%%%%%%%%%%%%%%%%%%%%%%%%%%

\subsection{Resonant breaking}\label{sec:condition_for_resonant_breaking}
Even when entering the feeding zone, planetesimals trapped in the mean motion resonances are not  always accreted by the migrating protoplanet.
The conjunction point of a trapped planetesimal with the protoplanet is given in terms of the mean longitude as: 
\begin{align}
     \lambda_{\rm con} = \varphi + \varpi,
\end{align}
where $\varpi$ is the longitude of pericenter. 
The resonant argument, $\varphi$, which is the displacement of the longitude of conjunction from the pericenter, takes $\varphi_{\rm eq} \sim 0$ during the converging orbital evolution.
This means that every conjunction occurs at the pericenter of the planetesimal orbit; namely, the planetesimal's orbit is far from the protoplanet's one at the conjunction.
Thus, trapped planetesimals cannot enter the Hill sphere of the protoplanet and planetesimal accretion does not occur \citep[see Figs.~C.1 and C.2 in][]{Shibata+2020}.

A planetesimal is captured by the migrating protoplanet only if the resonant trapping is  broken.
Under the forces damping and enhancing the planetesimal's eccentricity, the orbit of the trapped planetesimal becomes unstable and starts to oscillate.
Once the amplitude of oscillation exceeds the resonant width, the resonant trapping is broken.
This phenomenon is called the \textit{overstable libration} and \citet{Goldreich+2014} is derived the condition required for the instability to grow.
Substituting Eqs.~(\ref{eq:aerodynamic_damping_ecc}) and (\ref{eq:equilibrium_eccentricity_app}) into Eq.~(28) of \citet{Goldreich+2014}, we obtain the breaking condition for resonant trapping due to overstable libration under the aerodynamic gas drag as: 
\begin{align}
    3.29 \left(\frac{j}{j-1}\right)^{3/2} \frac{M_{\rm p}}{M_{\rm s}} < \frac{\tau_{\rm aero,0}}{\tau_{{\rm tide},a}}. \label{eq:overstable_libration_due_to_gasdrag}
\end{align}
Even once this condition is satisfied, it takes a time of $\sim\tau_{{\rm aero},e}$ for the instability to grow \citep{Goldreich+2014}.
If $\tau_{{\rm aero},e}~>~\tau_{{\rm tide},a}$, the trapped planetesimal is shepherded into the inner disk before escaping the resonance.
The second condition required for the overstable libration is $\tau_{{\rm aero},e}~<~\tau_{{\rm tide},a}$, which one can write as: 
\begin{align}
    \frac{\tau_{\rm aero,0}}{\tau_{{\rm tide},a}} < 0.54 (j-1)^{-1/2}, \label{eq:condition_for_resonant_breaking_grow}
\end{align}
using Eqs.~(\ref{eq:aerodynamic_damping_ecc}) and (\ref{eq:equilibrium_eccentricity_app}).

\subsection{Required condition for the Accretion Sweet Spot}\label{sec:condition_for_sweet_spot}
The required condition for the \SSP\ is summarised as: 
\begin{align}
    {\rm max} \left. 
        \begin{cases}
            1.54 (j-1) {e_{\rm cross}}^3\\
            \displaystyle{            
                3.29 \left(\frac{j}{j-1}\right)^{3/2} \frac{M_{\rm p}}{M_{\rm s}}
            }
        \end{cases}
    \right\}
    < \frac{\tau_{\rm aero,0}}{\tau_{{\rm tide},a}} < 0.54 (j-1)^{-1/2} \label{eq:sweet_spot}.
\end{align}
The inner edge of the \SSP\ is regulated by the aerodynamic shepherding and resonant shepherding.
The dominating  process depends on the planetary mass.
On the other hand, the outer edge of \SSP\ is regulated by the resonant shepherding only. 

To derive above equations, we assume that $\tau_{\mathrm{aero},a}\gg\tau_{\mathrm{tide},a}$.
Using the equilibrium eccentricity, we transform this condition as
\begin{align}
    \frac{2.18 {\eta_{\rm gas}}^{3/2}}{\sqrt{j-1}} \ll \frac{\tau_{\mathrm{aero},0}}{\tau_{\mathrm{tide},a}}.
\end{align}
In the nominal disk model, $\eta_{\rm gas}$ is the order of $10^{-3}$ and this condition is easily achieved in the SSP.
Thus, the eqs.~(\ref{eq:sweet_spot}) can be used for small planetesimals as long as the gas drag strength is scaled by eq.~(\ref{eq:aerodynamic_damping_timescale}).
We find that the location of the \SSP\ is determined by the ratio of the gas drag damping timescale $\tau_{\rm aero,0}$ and the planetary migration timescale $\tau_{{\rm tide},a}$.
We therefore conclude that our theoretical prediction is robust (and rather general) since 
the derivation of this condition does not depend on the assumed disk and migration models.  
As a result, Eq.~(\ref{eq:sweet_spot}) can be generally applied to various models of planetesimal accretion.

\section{Comparison with numerical results}
\label{sec:sweet_spot_numerical}
In this section we validate the derived equation for the \SSP\ by comparison with numerical results.
To that end, we perform $N$-body simulations similar to those presented in \citet{Shibata+2020}.
We adopt different values of model parameters from those in \citet{Shibata+2020} in order to identify whether the \SSP\ is indeed regulated by the ratio of the two timescales, $\tau_{\rm aero,0}$ and $\tau_{{\rm tide},a}$.
%Note that we changed the disk model from the viscous evolving disk to the minimum mass solar nebulae \citep{Hayashi1981} for simplifying the simulation. 
%An overview of the model is given in Appendix~\ref{sec:app_model}.

\subsection{Numerical Model and Settings}\label{sec:model_settings}
We consider the following scenario: 
Gas accretion has terminated and the young giant planet no longer grows in mass. 
The planet then migrates radially inward from a given semi-major axis within a protoplanetary disk. 
Initially there are many single-sized planetesimals interior to the planet's orbit.  
The migrating planet then encounters these planetesimals and captures some of them. 
The planetesimals are represented by test particles and, therefore, are affected only by the gravitational forces from the central star and planet, and the drag force by the disk gas.
Here, we summarise our numerical model.
The details of our numerical model are described in Appendix~\ref{app:method}.

We assume that the planetary migration timescale is given by:  
\begin{align}
    \tau_{{\rm tide},a} &= \tau_{\rm tide,0} \left( \frac{a_{\rm p}}{1 \AU} \right)^{1/2}, \label{eq:type2_migration_timescale}
\end{align}
where $\tau_{\rm tide,0}$ is a scaling parameter.
For the drag force by the disk gas, we adapt a model by \citet{Adachi+1976}.
The gas drag force depends on the size of planetesimal $R_{\rm pl}$ and the profile of ambient disk gas.
To focus on the dynamic process, we adopt simply the minimum-mass solar nebula model \citep{Hayashi1981} as our gas disk model.
The surface density $\Sigma_{\rm gas}$ and the disk' temperature $T_{\rm disk}$ are given by: 
\begin{align}
    \Sigma_{\rm gas} &= \Sigma_{\rm gas,0} \left( \frac{r}{1\AU} \right)^{-\alpha_{\rm disk}}, \label{eq:Disk_Gas_MMSN} \\
    T_{\rm disk} &= T_{\rm disk,0} \left(\frac{r}{1\AU} \right)^{-2\beta_{\rm disk}}, \label{eq:Disk_Temperature_MMSN}
\end{align}
where $r$ is the radial distance from the initial mass centre of the star-planet system, $\Sigma_{\rm gas,0} = 1.7 \times 10^3 \g/\cm^2$, $\alpha_{\rm disk}=3/2$, $T_{\rm disk,0}=280~\K$ and $\beta_{\rm disk}=1/4$.
%The surface density $\Sigma_{\rm gas}$ is given by: 
%\begin{align}\label{eq:Disk_Gas_MMSN}
%    \Sigma_{\rm gas} = \Sigma_{\rm gas,0} \left( \frac{r}{1\AU} \right)^{-\alpha_{\rm disk}},
%\end{align}
%where $r$ is the radial distance from the initial mass centre of the star-planet system, $\Sigma_{\rm gas,0} = 1.7 \times 10^3 \g/\cm^2$ and $\alpha_{\rm disk}=3/2$.
%The disk' temperature $T_{\rm disk}$ is given by: 
%\begin{align}
%    T_{\rm disk} = T_{\rm disk,0} \left(\frac{r}{1\AU} \right)^{-2\beta_{\rm disk}},
%\end{align}
%where $T_{\rm disk,0}=280~\K$ and $\beta_{\rm disk}=1/4$.
The protoplanetary disk is assumed to be vertically isothermal.

For the planetesimals we adopt a simple surface density profile given by:
\begin{align}
    \Sigma_{\rm solid} = Z_{\rm s} \Sigma_{\rm gas,0} \left( \frac{r}{1\AU} \right)^{-\alpha_{\rm disk}^{\prime}}, \label{eq:Model_Disk_Solid_Surface_Density}
\end{align}
where $Z_{\rm s}$ is the solid-to-gas ratio (or the metallicity) and $\alpha_{\rm disk}^{\prime}=\alpha_{\rm disk}=3/2$.
%We assume that $Z_{\rm s}$ equals the metallicity of the central star.
To speed up the numerical integration, we follow the orbital motion of super-particles, each of which contains several equal-size planetesimals.
The super-particles are initially distributed in $a_{\rm pl,in}< r < a_{\rm pl,out}$ with $a_{\rm pl,in}=0.3\AU$ and $a_{\rm pl,out}=20~\AU$.
%The number of super-particles $N_{\rm sp}$ is $12 000$.
The planet is initially located in such a way that the inner boundary of the feeding zone is consistent with the outer edge of the planetesimals disk, namely: 
\begin{align}
    a_{\rm p,0} = \frac{a_{\rm pl,out}}{1-2\sqrt{3} h}. \label{eq:part1_initial_semimajoraxis_planet}
\end{align}

The calculation is artificially stopped once the planet reaches the orbit of $a_{\rm p,f}=0.5 \AU$.
First, we investigate the planetesimal accretion in a reference case, where $\tau_{{\rm tide},0}= 1 \times 10^5~\yr$, $R_{\rm pl}= 1 \times 10^7~\cm$ and $M_{\rm p} = 1 \times 10^{-3}M_{\rm s}$.
In the reference case, the total amount of planetesimals are $M_{\rm tot} \sim 43 \Mear$.
The choices of the parameter values for the reference model are summarised in Table~\ref{tb:settings}.
We perform a parameter study for various values of $\tau_{\rm tide,0}$, $R_{\rm pl}$, and $M_{\rm p}$.
The dynamical integration for the planetesimals is performed using the numerical simulation code developed by \citet{Shibata+2019}.
In this code, we integrate the equation of motion using the forth-order-Hermite integration scheme \citep{Makino+1992}.
For the timesteps, we adopt the method of \citet{Aarseth1985}.

\begin{table*}
	\centering
	\begin{tabular}{ llllll } % four columns, alignment for each
		%\hline
		%Parameters used in the model &&&&&\\
		\hline
		$M_{\rm s}$             & Mass of central star	                        & $1.0$				                & $\Msun$ 	\\
		$Z_{\rm s}$             & Metallicity of central star	                & $0.018$				            & - \\
		$M_{\rm p}$             & Mass of planet		                        & $1.0$		                        & $\Mjup$  \\
%		$R_{\rm p}$             & Radius of planet		                        & eq.~(\ref{eq:Rplanet_cap})        & \\
%		$\rho_{\rm p}$          & Mean density of planet                        & $0.125$                           & $\g~\cm^{-3}$ \\
        $a_{\rm p,0}$           & Initial semi-major axis of planet             & eq.~(\ref{eq:part1_initial_semimajoraxis_planet})  & \\
        $a_{\rm p,f}$           & Final semi-major axis of planet               & $0.5$                             & $\AU$ \\
		$\tau_{\rm tide,0}$     & Scaling factor of migration timescale	        & $1.0 \times 10^5$                 & $\yr$ \\
        $\Sigma_{\rm disk,0}$   & Surface density of disk gas at $1\AU$         & $1.7\times10^{3}$		            & $\g/\cm^2$  \\
        $T_{\rm disk,0}$        & Temperature at $1\AU$                         & $280$		                        & $\K$  \\
        $\alpha_{\rm disk}$     & Exponent of disk gas profile                  & $3/2$                             & - \\
        $\beta_{\rm disk}$      & Exponent of disk temperature profile          & $1/4$                             & - \\
        $R_{\rm pl}$            & Radius of planetesimal			            & $1.0 \times 10^7$	                & $\cm$		    \\
        $a_{\rm pl,in}$         & Inner edge of planetesimal disk	            & $0.3$	                            & $\AU$		    \\
        $a_{\rm pl,out}$        & Outer edge of planetesimal disk	            & $20$	                            & $\AU$		    \\
%        $\alpha_{\rm sp}$       & Exponent of test particle distribution        & $1$                               & - \\
		$\rho_{\rm pl}$         & Mean density of planetesimal			        & $2.0$	                            & $\g~\cm^{-3}$		    \\
		$N_{\rm sp}$            & Initial number of super-particles             & 12 000                            & - \\
		$C_{\rm d}$             & Non-dimensional drag coefficient              & 1                                 & - \\
		\hline
	\end{tabular}
	\caption{
	Parameters used in the reference model.
    }
    \label{tb:settings}
\end{table*}

\subsection{Comparison with the numerical result}\label{sec:sweet_spot_compare}
%%%%%%%%%%%%%%%%%%%%%%%%%%%%%%%%%%%%%%%%%%%%%%%%%%%%%%%%%%%%%
\begin{figure}
  \begin{center}
    \includegraphics[width=80mm]{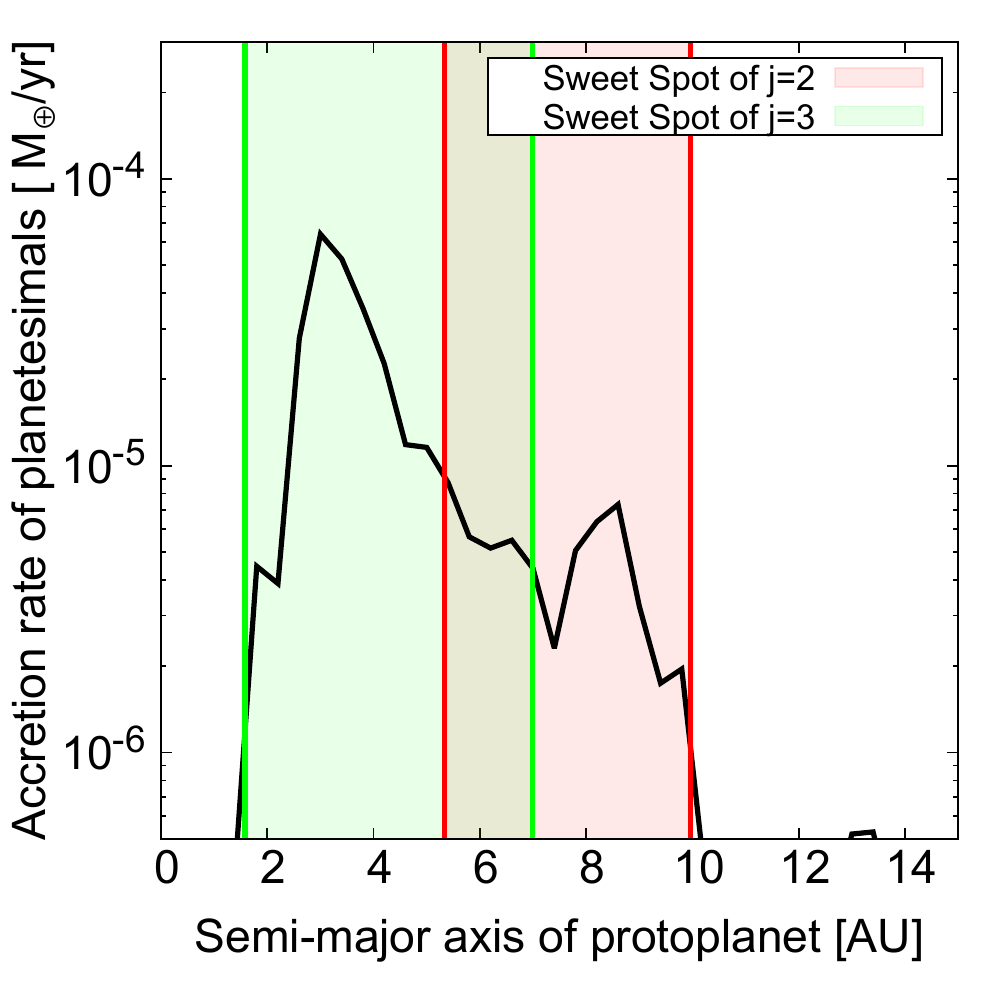}
    \caption{
    The planetesimal accretion rate vs.~the semi-major axis of the migrating planet. 
    The vertical solid lines show the boundaries of the sweet spot for planetesimal accretion (\SSP) given by Eq.~(\ref{eq:sweet_spot}).
    The red lines are for the case of $j=2$ and the green lines are for $j=3$.
    The filled areas show the \SSP.
    }
    \label{fig:Axi_dMcap}
  \end{center}
\end{figure}
%%%%%%%%%%%%%%%%%%%%%%%%%%%%%%%%%%%%%%%%%%%%%%%%%%%%%%%%%%%%%

Figure~\ref{fig:Axi_dMcap} shows the planetesimal accretion rate at each semi-major axis of the migrating protoplanet, obtained in the simulation for the reference model.
Planetesimal accretion is found to occur efficiently in a region of $2~\AU \lesssim a_{\rm p} \lesssim 10~\AU$. 
Beyond 10~$\AU$, planetesimals are trapped in the mean motion resonances of $j$ = 2 and 3 with the migrating protoplanet.
%When the protoplanet migrates in $\gtrsim10\AU$, 
%$j=2$ and $j=3$ mean motion resonances trap planetesimals as 
%the protoplanet migrates. 
Our analysis of the numerical results shows  that the condition of overstable libration is achieved, but the timescale of the instability growth is longer than the migration timescale; thus, the trapped planetesimals are shepherded without being captured by the protoplanet.
Around $\sim10\AU$, planetesimals trapped in the $2:1$ mean motion resonance start to escape due to the overstable libration.
The large amount of shepherded planetesimals are supplied into the planetary feeding zone, so that the accretion rate peaks at around $\sim8\AU$.
Aerodynamic shepherding starts around $5\AU$ for the planetesimals in the $2:1$ mean motion resonance, halting the supply of planetesimals through the $2:1$ mean motion resonance and reducing the accretion rate significantly.
Instead, planetesimal supply resumes through the $3:2$ mean motion resonance around $7\AU$ and the accretion rate increases again.
More specifically, planetesimals that remain to be brought into the feeding zone through $2:1$ mean motion resonance move to the $3:2$ resonance; namely, the accretion band shifts from $j=2$ to $j=3$. 
The accretion rate peaks again around $3\AU$ and then rapidly decreases around $2\AU$ due to aerodynamic shepherding.

In the simulation setting, the timescale ratio is given by: 
\begin{align}
    \frac{\tau_{\rm aero,0}}{\tau_{{\rm tide},a}} &= \left| \frac{\tau_{\rm aero,0}}{\tau_{{\rm tide},a}} \right|_{1 \AU} \left( \frac{j-1}{j} \right)^{1/3} a^{11/4}, \label{eq:tau_frac_dependence_on_a} \\
                                                &= \left| \frac{\tau_{\rm aero,0}}{\tau_{{\rm tide},a}} \right|_{1 \AU} \left( \frac{j-1}{j} \right)^{13/6} {a_{\rm p}}^{11/4}, \label{eq:tau_frac_dependence_on_ap} 
\end{align}
where
\begin{align}
    \left| \frac{\tau_{\rm aero,0}}{\tau_{{\rm tide},a}} \right|_{1 \AU} = 4.2 \times 10^{-3} \left(\frac{R_{\rm pl}}{10^7~\cm} \right) \left(\frac{\tau_{\rm tide,0}}{10^5~\yr} \right)^{-1}. \label{eq:tau_frac_0}
\end{align}
The red-shaded and green-shaded areas in Fig.~\ref{fig:Axi_dMcap} show the predicted regions of the \SSP, %given by Eq.~(\ref{eq:sweet_spot}), (\ref{eq:tau_frac_dependence_on_ap}) and (\ref{eq:tau_frac_0}), where 
$5~\AU~\lesssim~a_{\rm p}~\lesssim~10~\AU$ for $j=2$ and $1~\AU~\lesssim~a_{\rm p}~\lesssim~7~\AU$ for $j=3$, respectively, which one obtains using Eqs.~(\ref{eq:tau_frac_dependence_on_ap}) and (\ref{eq:tau_frac_0}) in Eq.~(\ref{eq:sweet_spot}).
Those regions turn out to cover the areas where the accretion rate is higher than  $10^{-6}~\Mear/\yr$.
Thus, we conclude that the derived equation for the \SSP\ reproduces well the numerical results in the reference model.

\subsection{Results of parameter studies}
As seen in Eq.~(\ref{eq:sweet_spot}), the location of the \SSP\ depends on the ratio of the gas drag damping timescale $\tau_{{\rm aero}, 0}$ to the planetary migration timescale $\tau_{{\rm tide}, a}$ and the planetary mass $M_{\rm p}$.
To investigate this dependence, we perform parameter studies for the size of planetesimals $R_{\rm pl}$, which determines $\tau_{\mathrm{aero}, 0}$, the scaling factor of migration timescale $\tau_{\rm tide,0}$, and the planetary mass $M_{\rm p}$.

\subsubsection{Dependence on the planetesimal's  size}
%%%%%%%%%%%%%%%%%%%%%%%%%%%%%%%%%%%%%%%%%%%%%%%%%%%%%%%%%%%%%
\begin{figure}
  \begin{center}
    \includegraphics[width=80mm]{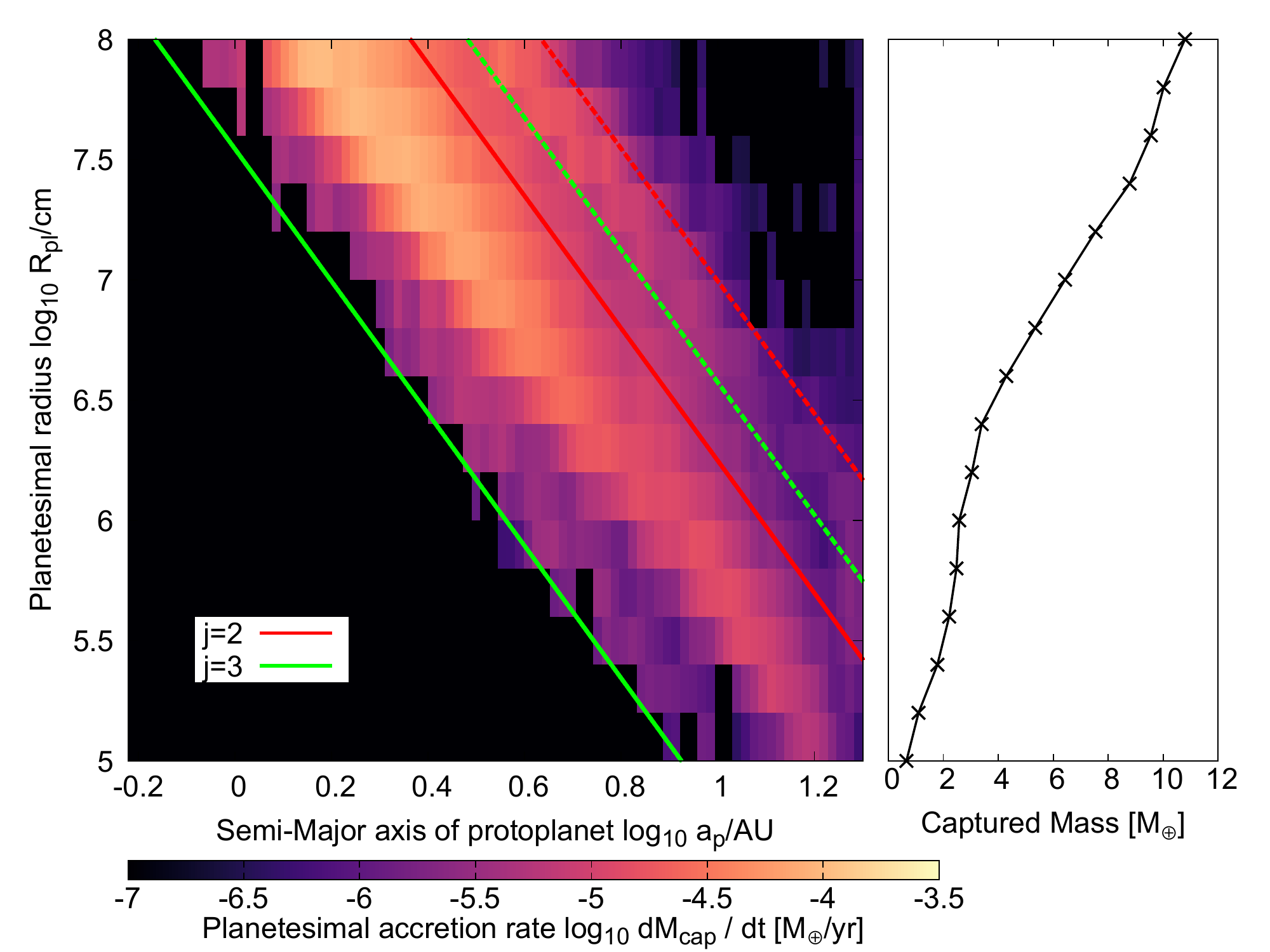}
    \caption{
    The results of the parameter study regarding the size of planetesimals $R_{\rm pl}$.
    {\bf Left:} Planetesimal accretion rate shown as a color contour on a plane of the semi-major axis of the migrating protoplanet and the planetesimal's radius.
    The analytically predicted inner and outer boundaries of the sweet spot are indicated with solid and dashed lines, respectively.
    The red lines are for $j=2$ and the green lines are for $j=3$.
    {\bf Right:} Total mass of captured planetesimals as a function of the planetesimal radius.
    }
    \label{fig:Result_ParameterStudy_Rpl}
  \end{center}
\end{figure}
%%%%%%%%%%%%%%%%%%%%%%%%%%%%%%%%%%%%%%%%%%%%%%%%%%%%%%%%%%%%%
Since the gas drag damping timescale $\tau_{\rm aero,0}$ increases with the size of planetesimals, the location of the \SSP\ is predicted to be closer to the central star for larger planetesimals (see, e.g., Eqs.~(\ref{eq:tau_frac_dependence_on_ap}) and (\ref{eq:tau_frac_0})).
We perform the numerical simulations, changing the size of planetesimals from $1 \times 10^5~\cm$ to $1 \times 10^8~\cm$ (or $\sim0.01\Mear$). 
Note that the assumption that mutual gravitational interaction among  planetesimals is negligible is not valid for large planetesimals with sizes of $R_{\rm pl} \sim 10^8~\cm$.
Nevertheless, a study for a wide parameter range is useful to deepen the  physical understanding of the effect of the damping timescale on planetesimal accretion.

Figure~\ref{fig:Result_ParameterStudy_Rpl} shows the results of the parameter study when changing the planetesimal's radius $R_{\rm pl}$.
In the left panel, the planetesimal accretion rate is shown as a color contour on a plane of the semi-major axis of the migrating protoplanet and the planetesimal's radius.
The analytically predicted inner and outer boundaries of the \SSP\ derived in sec.~\ref{sec:orbital_evolution_in_resonant_trapping} are shown with the solid and dashed lines, respectively.
The red lines are for $j=2$ and the green lines are for $j=3$.
In the right panel, the total mass of captured planetesimals $M_{\rm cap, tot}$ is plotted as a function of the planetesimal's radius.

The location of the \SSP\ (i.e., the region of high accretion rates) is found to be well reproduced by the analytical expressions derived in sec.~\ref{sec:orbital_evolution_in_resonant_trapping}.
The maximum accretion rate and the total captured heavy-element mass are higher for larger planetesimals.
This is because as the \SSP\ moves inward, the amount of planetesimals shepherded into the \SSP\ increases.
More planetesimals are shepherded into the \SSP\ when the planetesimals are larger, where the \SSP\ locates farther from the initial planet location (or closer to the central star).

\subsubsection{Dependence on the planetary migration timescale}
%%%%%%%%%%%%%%%%%%%%%%%%%%%%%%%%%%%%%%%%%%%%%%%%%%%%%%%%%%%%%
\begin{figure}
  \begin{center}
    \includegraphics[width=80mm]{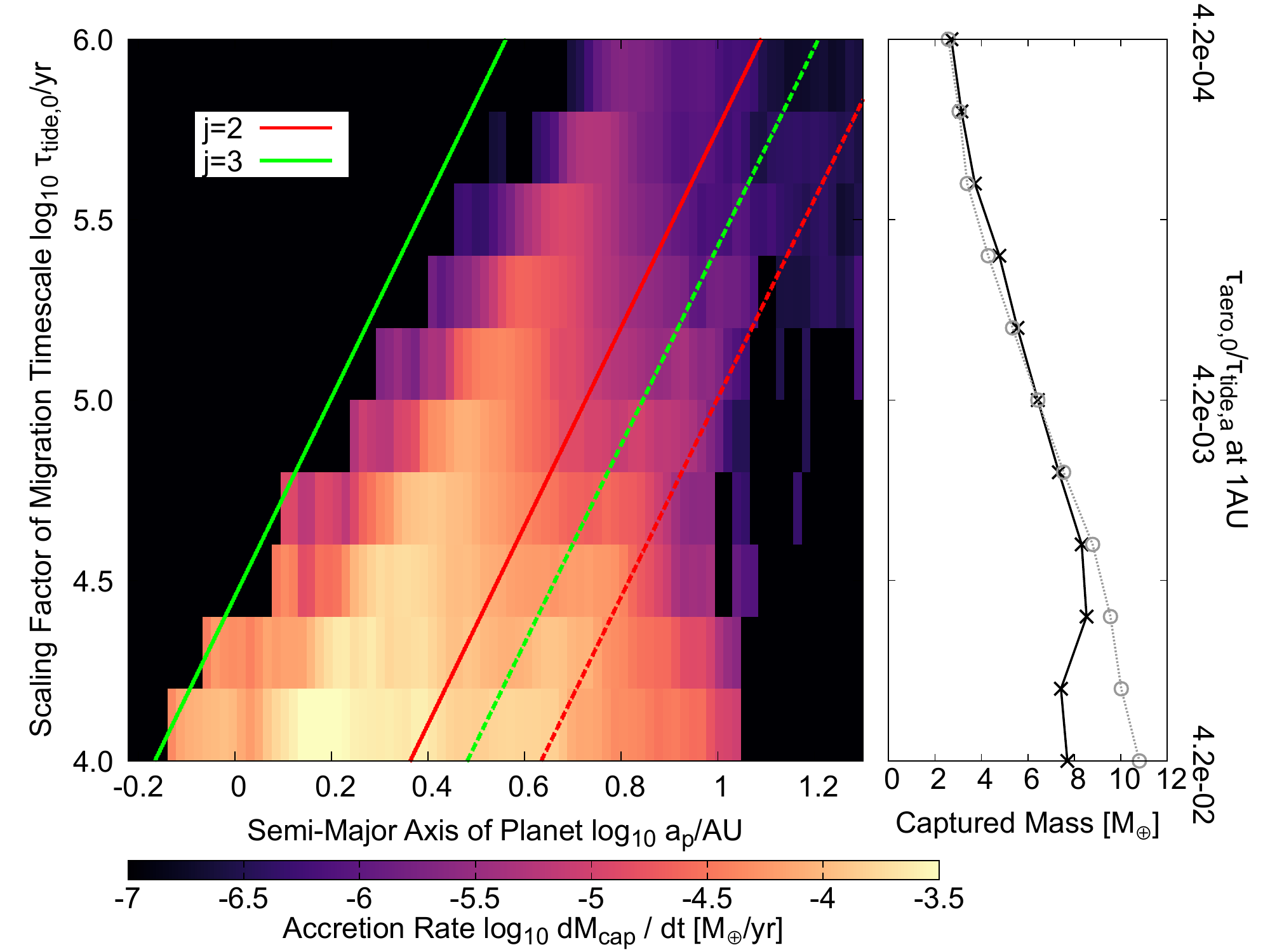}
    \caption{
    Same as Fig.~\ref{fig:Result_ParameterStudy_Rpl}, but for the results of the parameter study regarding the migration timescale $\tau_{\rm tide,0}$.
    The gray circles in the right panel show the results of the parameter study regarding the size of planetesimals $R_{\rm pl}$ for comparison using the fraction of timescales $\tau_{\mathrm{aero}, 0}/\tau_{\mathrm{tide}, a}$ in vertical axis.
    }
    \label{fig:Result_ParameterStudy_Tmig0}
  \end{center}
\end{figure}
%%%%%%%%%%%%%%%%%%%%%%%%%%%%%%%%%%%%%%%%%%%%%%%%%%%%%%%%%%%%%
Next, we perform a parameter study when changing the planetary migration timescale. 
Our analytic consideration above indicates that the location of the \SSP\ is closer to the central star for faster planetary migration. 
We perform numerical simulations with various migration timescales by changing the scaling factor of migration timescale $\tau_{\rm tide,0}$ from $1 \times 10^4~\yr$ to $1 \times 10^6~\yr$.

Figure~\ref{fig:Result_ParameterStudy_Tmig0} shows the dependence of the results to the migration timescale $\tau_{\rm tide,0}$.
Indeed, in the left panel, the \SSP\ is found to be closer to the star for faster planetary migration (or smaller $\tau_{\rm tide,0}$). 
Since the \SSP\, moves inward as $\tau_{\rm tide,0}$ decreases, for the same reason described above, the total mass of captured planetesimals $M_{\rm cap, tot}$ increases with decreasing $\tau_{\rm tide,0}$, which is confirmed in the right panel.
The right axis shows $\tau_{\mathrm{aero}, 0}/\tau_{\mathrm{tide}, a}$
at $1\AU$ calculated with Eq.~(\ref{eq:tau_frac_0}).
For comparison, we also plot with grey circles $M_{\rm cap, tot}$ vs. $\tau_{\mathrm{aero}, 0}/\tau_{\mathrm{tide}, a}$ relation by converting the result shown in the right panel of Fig.~\ref{fig:Result_ParameterStudy_Rpl} when using  Eq.~(\ref{eq:tau_frac_0}).
Except for $\tau_{\rm tide,0}$ $\lesssim$ $10^{4.6}\yr$, the two results are similar. 
This confirms that the location of the \SSP\ scales with the ratio of the timescales $\tau_{\rm aero,0}/\tau_{\rm tide,a}$.

The effects of planetary migration, however, seems somewhat more complicated relative to the dependence on the planetesimal's radius.
First, as shown in the left panel, for relatively fast migration ($\tau_{\rm tide,0} \lesssim 10^{4.6}$~yr), the outer boundary of the \SSP\, expands beyond the one inferred analytically. 
Second, the $M_{\rm cap, tot}$ vs. $\tau_{\mathrm{aero}, 0}/\tau_{\mathrm{tide}, a}$ relation differs between the cases when we vary $R_{\rm pl}$ and varying  $\tau_{\mathrm{tide},0}$ for $\tau_{\rm tide,0} \lesssim 10^{4.6}~\yr$. 
This is related to the stability of resonant trapping.
As shown in sec.~\ref{sec:condition_for_resonant_trapping}, the condition required for resonant trapping is that the timescale for planetesimals to cross the mean motion resonance is longer than the libration timescale.
Using Eq.~(\ref{eq:condition_for_resonant_trapping_mig}), we find that stable resonant trapping is difficult for fast planetary migration like $\tau_{\rm tide,0} \lesssim 10^4~\yr$.
%In this case, the planetesimals are barely trapped into the resonances, but the resonant trapping is unstable due to the fast planetary migration.

\begin{figure}
  \begin{center}
    \includegraphics[width=80mm]{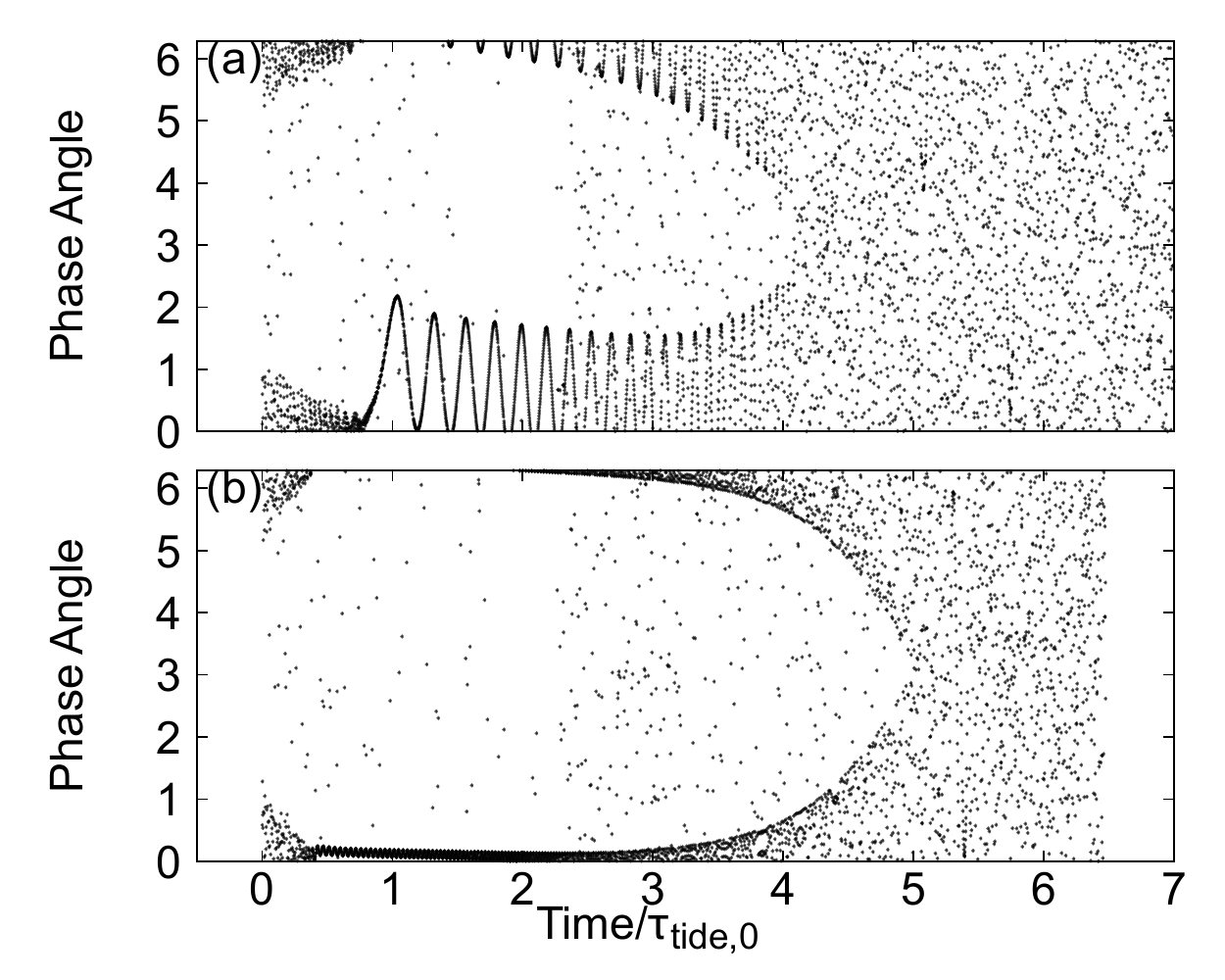}
    \caption{
    Change in the phase angle of a planetesimal initially located at $a_{\rm 0}=13.8~\AU$.
    Panel~(a) shows the case of $\tau_{\rm tide,0} = 1 \times 10^4~\yr$ and panel~(b) shows the case of $\tau_{\rm tide,0} = 1 \times 10^5~\yr$. 
    In panel~(a), the phase angle oscillates indicating that the resonant trapping is unstable due to the fast planetary migration.
    Then, the resonant trapping is broken earlier in the case of panel~(a) than that shown in panel~(b). 
    }
    \label{fig:Result_ParameterStudy_time_phase_angle}
  \end{center}
\end{figure}

Figure~\ref{fig:Result_ParameterStudy_time_phase_angle} shows the changes in the phase angle of a planetesimal initially located at $a_{\rm 0}=13.8~\AU$ in the cases of (a) $\tau_{\rm tide,0}=10^4~\yr$ and (b) $\tau_{\rm tide,0}=10^5~\yr$.
In panel (a), the phase angle continues libration even after trapped into the resonance.
This is because fast planetary migration perturbs the orbit of the trapped planetesimal and accelerates the overstable libration. 
The resonant trapping is, thus, broken earlier in the case of $\tau_{\rm tide,0}=10^4~\yr$ than in the case of $\tau_{\rm tide,0}=10^5~\yr$, which results in the outward expansion of the \SSP.
However, such an outward shift of the outer edge of the \SSP\ does not lead to an  increase in the mass of captured planetesimals.
This is because fast planetary migration breaks the resonant trapping as soon as the trapped planetesimal reaches the equilibrium condition.
The planetesimal in panel (a) escapes from the resonant trapping with $e\sim0.7$ and that in panel (b) with $e\sim0.4$.
The capture probability of planetesimals decreases with increasing eccentricity \citep{Ida+1989}; thus, the total mass of captured planetesimals does not increase with the expansion of the \SSP\ in the cases of $\tau_{\rm tide,0}\lesssim10^{4.6}~\yr$.

%%%%%%%%%%%%%%%%%%%%%%%%%%%%%%%%%%%%%%%%%%%%%%%%%%%%%%%%%%%%%
%%%%%%%%%%%%%%%%%%%%%%%%%%%%%%%%%%%%%%%%%%%%%%%%%%%%%%%%%%%%%

\subsubsection{Dependence on the planetary mass}
%%%%%%%%%%%%%%%%%%%%%%%%%%%%%%%%%%%%%%%%%%%%%%%%%%%%%%%%%%%%%
\begin{figure}
  \begin{center}
    \includegraphics[width=80mm]{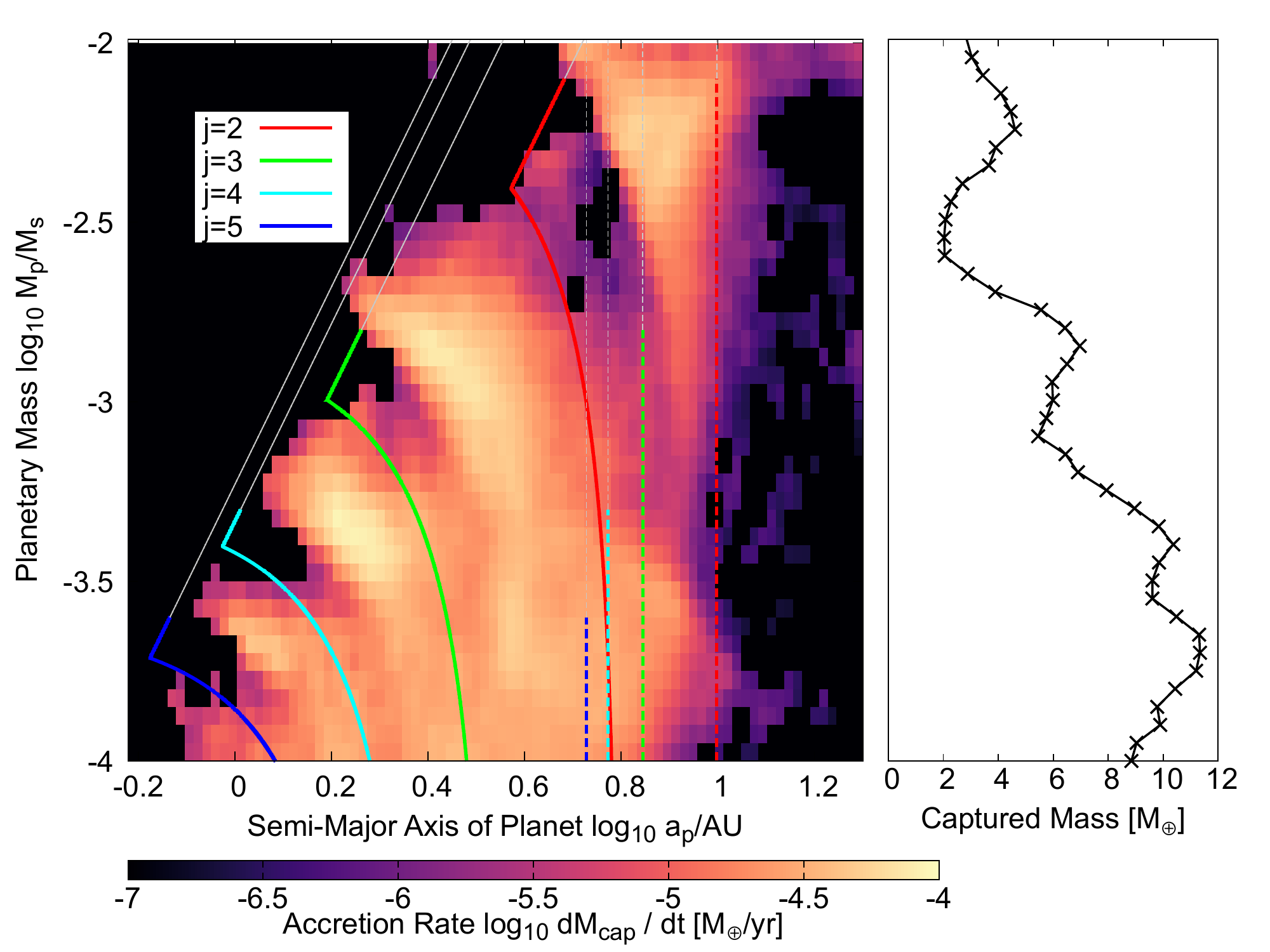}
    \caption{
    Same as Fig.~\ref{fig:Result_ParameterStudy_Rpl}, but for the results of the parameter study when varying the protoplanet's mass $M_{\rm p}$.
    The red, green, skyblue, and blue lines represent the \SSP\ inner (solid lines) and outer (dashed lines) boundaries determined by the resonant trapping, which is predicted by Eq.~(\ref{eq:sweet_spot}) for $j$ = 2, 3, 4, and 5, respectively.
    The coloured lines change into the gray lines once the protoplanet mass exceeds a threshold value, beyond which the accretion bands disappear (see Fig.~\ref{fig:Mp_EccCross}).
    }
    \label{fig:Result_ParameterStudy_Mp}
  \end{center}
\end{figure}
%%%%%%%%%%%%%%%%%%%%%%%%%%%%%%%%%%%%%%%%%%%%%%%%%%%%%%%%%%%%%
Finally, we perform a parameter study of the simulations when varying the mass of the migrating protoplanet $M_{\rm p}$ in the range of $M_{\rm p}/M_{\rm s}$ between $1 \times 10^{-4}$ and $1 \times 10^{-2}$.
The results are summarised in Figure~\ref{fig:Result_ParameterStudy_Mp}. 
We find that the semi-major axis of the outer boundary of the \SSP\, is insensitive to the protoplanet mass, which is consistent with Eq~(\ref{eq:sweet_spot}). 
In this figure we indicate the locations of the inner edge of the \SSP\, for $j$ = 4 and 5 in addition to those for $j$ = 2 and 3 given by Eq.~(\ref{eq:sweet_spot}), confirming that Eq.~(\ref{eq:sweet_spot}) accuratly reproduces the numerical result.
As shown in Fig.~\ref{fig:Mp_EccCross}, the crossing points between the resonance centre and the feeding zone boundary disappear once the planetary mass exceeds a specific value. 
In Fig.~\ref{fig:Result_ParameterStudy_Mp}, we show the $M_{\rm p}$-$a_{\rm p}$ relation corresponding to the disappearance of the crossing points with gray solid lines.
This is also consistent with the numerical result.

As found in \citet{Shibata+2020}, the total mass of captured planetesimals $M_{\rm cap,tot}$ changes with the planetary mass in a non-monotonic manner and takes local maxima at $M_{\rm p}/M_{\rm s}\sim10^{-3.6}$, $10^{-3.3}$, $10^{-2.7}$ and $10^{-2.2}$. 
This periodical changes come from the corresponding shifts of the inner boundary of the  \SSP.
The shepherding process that regulates the inner boundary of the \SSP\ changes from the aerodynamic shepherding to the resonant shepherding with increasing planetary mass.  
The accretion bands also change with the protoplanet mass.
%This change of the regulating process occurs at $M_{\rm p}/M_{\rm s}\sim10^{-3.6}$ for $j=5$, $M_{\rm p}/M_{\rm s}\sim10^{-3.4}$ for $j=4$, $M_{\rm p}/M_{\rm s}\sim10^{-3.0}$ for $j=3$ and $M_{\rm p}/M_{\rm s}\sim10^{-2.4}$ for $j=2$.
%Due to 
Because of these effects, the location of the inner boundary of the \SSP\ shifts  non-monotonically which results in the non-monotonic change of $M_{\rm cap,tot}$. %the total captured mass of planetesimals .
The global decreasing trend with increasing protoplanet mass comes from the outward shift of the inner boundary.

\section{Discussion}
\label{sec:discussion}
%\subsection{Comparison with \citet{Tanaka+1999}}
\subsection{Collision and ablation of highly-excited planetesimal}
\label{sec:Discussion_Collisions}
Our model neglects the effect of planetesimal collisions.
During the planetary migration, planetesimals are swept by the resonances and the enhanced local density of planetesimals increases the frequency of planetesimal collisions, which might initiate the collisional cascade \citep{Batygin+2015}.
Collisional cascade \citep[e.g.][]{Dohnanyi+1969} changes the size distribution of planetesimals and the perturbation adding on the planetesimal orbits by the collisions would also affect the resonant configuration \citep{Malhotra1993b}.
In this section, we consider the effect of planetesimal collisions on the location of the SSP and the planetesimal accretion.

\subsubsection{Collision timescale of shepherded planetesimals}
%%%%%%%%%%%%%%%%%%%%%%%%%%%%%%%%%%%%%%%%%%%%%%%%%%%%%%%%%%%%%
\begin{figure}
  \begin{center}
    \includegraphics[width=80mm]{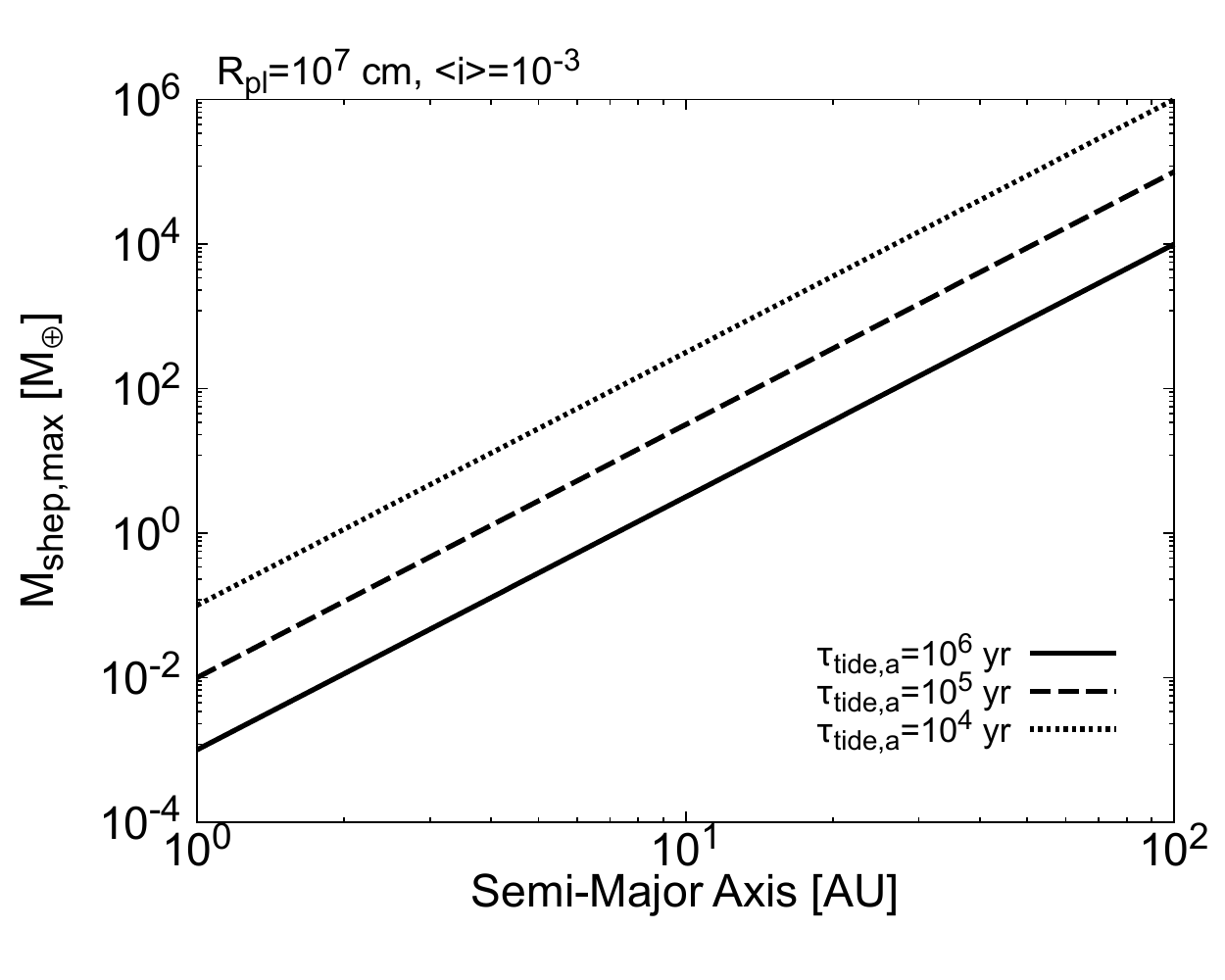}
    \caption{
    Maximum mass of planetesimals that can be shepherded in the mean motion resonances without effective mutual collisions.
    We show the cases of $R_{\rm pl}=10^7 \cm$ and $\left< i \right>=10^{-3}$.
    The solid, dashed, and dotted lines correspond to  $\tau_{\mathrm{tide,}a}=10^6 \yr$, $10^5 \yr$ and $10^4 \yr$, respectively.
    }
    \label{fig:Result_Discussion_Mshep}
  \end{center}
\end{figure}
%%%%%%%%%%%%%%%%%%%%%%%%%%%%%%%%%%%%%%%%%%%%%%%%%%%%%%%%%%%%%
First, we estimate the collision timescale using the number density of planetesimals $n_{\rm pl}$, the collision cross section $\sigma$ and the relative velocity $u$.
We consider a swarm of planetesimals that is trapped in mean motion resonances with total mass of $M_{\rm shep}$.
For simplicity, we consider that these planetesimals have the same mass.
The planetesimals are distributed in the ring-like region with a  width of $\sim 2 a\left< e\right>$ and thickness of $\sim 2 a \tan \left< i\right>$.
The collision timescale between these planetesimals $\tau_{\rm col}$ is given as: 
\begin{align}
    \tau_{\rm col}  &= \left( n_{\rm pl} \sigma u \right)^{-1}, \\
                    &= \left(\frac{M_{\rm shep}}{2 \pi a \cdot 2 a \left<e\right> \cdot 2 a \tan \left<i\right>} \cdot \pi {R_{\rm pl}}^2 \cdot \left< e\right> v_{\rm K} \right)^{-1}, \\
                    &= 3\times10^5 \yr \left( \frac{M_{\rm shep}}{10 M_{\oplus}} \right)^{-1} \left( \frac{\rho_{\rm pl}}{2 \g/\cm^3} \right) \left( \frac{R_{\rm pl}}{10^{7}\cm} \right)  \nonumber \\
                    & \left( \frac{a}{10 \AU} \right)^{7/2} \left( \frac{\tan \left< i\right>}{10^{-3}} \right) \left( \frac{M_{\rm s}}{M_{\odot}} \right)^{-1/2}.  \label{eq:tau_collision}
\end{align}
Here, we neglect the effect of gravitational focusing because of the high relative velocities of planetesimals.
$M_{\rm shep}$ increases as the protoplanet migrates and sweeps planetesimals, but decreases after reaching the SSP because the trapped planetesimals start to escape from the mean motion resonances.
As suggested by eq.~(\ref{eq:tau_collision}), $\tau_{\rm col}$ rapidly decreases with the inward planetary migration.
Planetesimals start to collide with each other where the collision timescale $\tau_{\rm col}$ is comparable to the migration timescale $\tau_{\mathrm{tide},a}$.
The maximum mass of planetesimals shepherded into the SSP without active planetesimal collisions $M_{\rm shep,max}$ is estimated as:
\begin{align}
    M_{\rm shep,max} &= 30 M_{\oplus} \left( \frac{\tau_{\mathrm{tide},a}}{10^5 \yr} \right)^{-1} \left( \frac{\rho_{\rm pl}}{2 \g/\cm^3} \right) \left( \frac{R_{\rm pl}}{10^7 \cm} \right) \nonumber \\ 
     &\left( \frac{a}{10 \AU} \right)^{7/2} \left( \frac{\tan \left< i\right>}{10^{-3}} \right) \left( \frac{M_{\rm s}}{M_{\odot}} \right)^{-1/2}. \label{eq:maximum_Mshep}
\end{align}
Figure~\ref{fig:Result_Discussion_Mshep} shows $M_{\rm shep,max}$ as a function of the radial distance from the central star.
During the planetary migration, despite the increase of $M_{\rm shep}$, $M_{\rm shep,max}$ rapidly decreases.
We conclude that the trapped planetesimals collide with each other during the planetary migration phase if the protoplanet migrates a long distance in the radial direction before entering the SSP.

\subsubsection{Collisional cascade in the resonant shepherding}
The outcome of planetesimal collision is determined by the specific energy, which is the kinetic energy of impacting planetesimal normalised by the mass of impacted planetesimal, given by \citep[e.g.][]{Armitage+2010}: 
\begin{align}\label{eq:specific_value}
    Q &= 1.1\times10^{12} {\rm erg}/{\g} \left( \frac{e}{0.5} \right)^2 \left( \frac{R_{\rm im}}{R_{\rm pl}} \right)^{3} \left(\frac{M_{\rm s}}{M_{\odot}} \right) \left( \frac{a}{10 \AU} \right)^{-1},
\end{align}
where $R_{\rm im}$ is the radius of the impacting planetesimal. %and $R_{\rm tg}$ are
The critical value for shattering planetesimals can be scaled as: 
\begin{align}
    Q_{\rm D}^{*} = q_{\rm s} \left( \frac{R_{\rm pl}}{1 \cm} \right)^{a} + q_{\rm g} \rho_{\rm pl} \left( \frac{R_{\rm pl}}{1 \cm} \right)^{b}. \label{eq:specific_value_dispersing}
\end{align}
\citet{Benz+1999} and \cite{Leinhardt+2009} determined the values of fitting parameters in this equation.
For icy planetesimals with an impact velocity of $3\km/\s$, they obtained $q_{\rm s}=1.6\times10^7~\erg/\g$, $q_{\rm g}=1.2~\erg~\cm^3/\g^2$, $a=-0.39$ and $b=1.26$.
Using eq.~(\ref{eq:specific_value}) and (\ref{eq:specific_value_dispersing}), we find that 
km-sized planetesimals are shattered by the collisions of planetesimals larger than $R_{\rm im,th}$, which is given as: 
\begin{align}\label{eq:Rbreak_threathhold}
    \frac{R_{\rm im,th}}{R_{\rm pl}} = 0.13 \left( \frac{R_{\rm pl}}{10^7 \cm} \right)^{0.42} \left( \frac{e}{0.5} \right)^{-2/3} \left(\frac{M_{\rm s}}{M_{\odot}} \right)^{-1/3} \left( \frac{a}{10 \AU} \right)^{1/3}.
\end{align}

We find that collisions of trapped planetesimals easily result in the dispersal and formation of many smaller planetesimals. 
The generated smaller planetesimals shortens the collisional timescale.
Namely, during the planetary migration phase, $M_{\rm shep}$ increases and once it exceeds the threshold value $M_{\rm shep,max}$, the trapped planetesimals start to collide with each other and initiate collisional cascade.
The planetesimal size distribution is largely changed and the major source of solid material would be the smaller planetesimals. % than $R_{\rm pl}$.
As shown by our simulations, the decrease in planetesimal size shifts the location of the SSP outward.
Therefore the possibility of collisional cascade occuring during planetary migration could affect the efficiency of planetesimal accretion.

The value of $M_{\rm shep,max}$ after the initiation of the collisional cascade would be affected by the amount of dispersed planetesimals left in the mean motion resonances. %larger than $R_{\rm im,th}$
Non-dispersed planetesimals with radii of $R_{\rm pl}$ can be supplied as the planet migrates, however,
if a large amount of dispersed planetesimals larger than $R_{\rm im,th}$ are left in the resonance, $M_{\rm shep,max}$ is decreased by a factor of $\sim{R_{\rm im,th}}/{R_{\rm pl}}$ at most.
As we discuss below, impacted planetesimals might be removed from the mean motion resonances and small fragments generated by the collisional cascades would be drifted into the inner disk region faster than the planet.
Thus, after the initiation of the collisional cascade, the size distribution of planetesimals in the resonant shepherding would be affected by various physical processes.
{\bf 
We suggest that the mutual collisions of planetesimals in the resonant shepherding could be important for determining the population of small objects in planetary systems. 
}

In the above discussion, we assumed that the relative velocity of planetesimals $u$ is approximated by $\sim~e~v_{\rm K}$.
However, the alignment of planetesimal orbits via gravitational perturbation from the protoplanet is known to reduce the relative velocity of planetesimals.
\citet{Guo+2021} investigated the effect of orbital alignment due to the secular perturbation which converges the longitude of pericenter into a certain value; namely, planetesimal orbits are aligned on the inertial plane.
Using N-body simulations, they found that the relative velocity of planetesimals are reduced outside the mean motion resonances.
According to their results, orbital alignment is not effective inside the mean motion resonances because of the resonant perturbation.
In the planetary migration phase, however, the resonance angle $\varphi$ {\bf converges} into zero and planetesimal orbits can be aligned on the plane rotating with the protoplanet (see fig.~C1 and C2 in \citet{Shibata+2020}).
In the ideal case, the relative velocity of these planetesimals is smaller than $\sim~e~v_{\rm K}$ and some collisions would result in sticking.
In addition, the mean inclination of planetesimals $\left< i \right>$ is difficult to estimate, but it is important for collision timescale. 
The viscous stirring between trapped planetesimals might affect the mean inclination of planetesimals $\left< i \right>$, which is not considered in our simulations.
Under effective viscous stirring, the velocity dispersion of the planetesimal swarm keeps a simple relation of $\left< e^2 \right>^{1/2}\sim 2 \left< i^2 \right>^{1/2}$ \citep[e.g.][]{Ohtsuki+2002}.
In this case, the collision timescale  increases and $M_{\rm shep,max}$ would be larger than indicated in fig.~\ref{fig:Result_Discussion_Mshep}.
However, it is still unclear whether the energy equipartition works in the resonant trapping where the planetesimal's eccentricity is excited by the protoplanet.
It is clear that  both $M_{\rm shep,max}$ and the planetesimal size distribution depends on the interactions between planetesimals in the mean motion resonances. We suggest that this topic should be investigated in detail in future work.

\subsubsection{Breakup of resonant trap}\label{sec:Discussion_CompareTI99}
%%%%%%%%%%%%%%%%%%%%%%%%%%%%%%%%%%%%%%%%%%%%%%%%%%%%%%%%%%%%%
\begin{figure}
  \begin{center}
    \includegraphics[width=80mm]{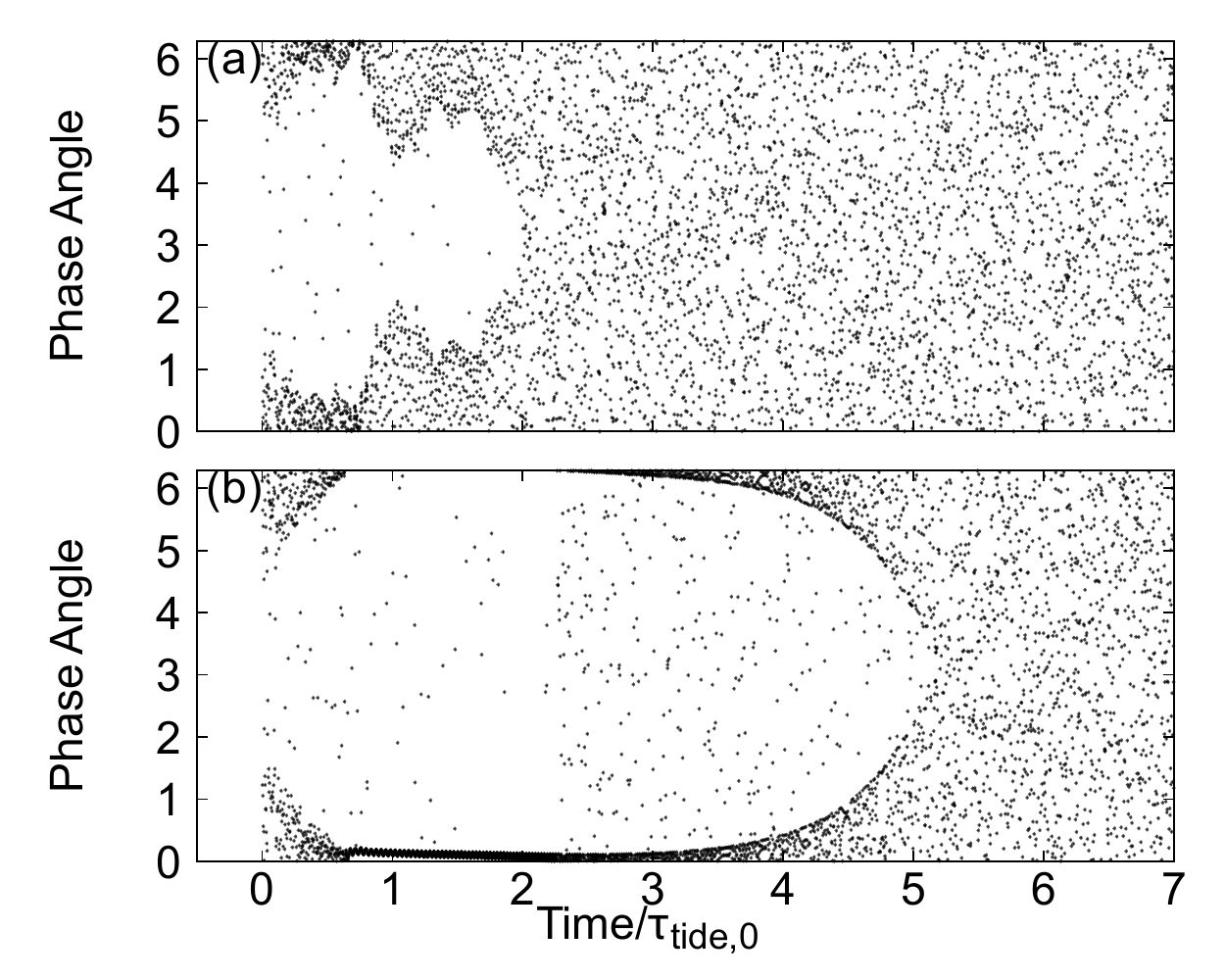}
    \caption{
    Same as Fig.~\ref{fig:Result_ParameterStudy_time_phase_angle}, but (a) for the case with artificial perturbation given by Eq.~(\ref{eq:artificial_perturbation}) and (b) without artificial perturbation.
    Here we use our reference model with $R_{\rm pl}=1\times10^7\cm$, $\tau_{\rm tide,0}=1\times10^5\yr$ and $M_{\rm p}/M_{\rm s}=1\times10^{-3}$.
    %\ikoma{FOR THE REFERENCE MODEL?}
    }
    \label{fig:Result_Discussion_time_phase_angle_wfp}
  \end{center}
\end{figure}

\begin{figure}
  \begin{center}
    \includegraphics[width=80mm]{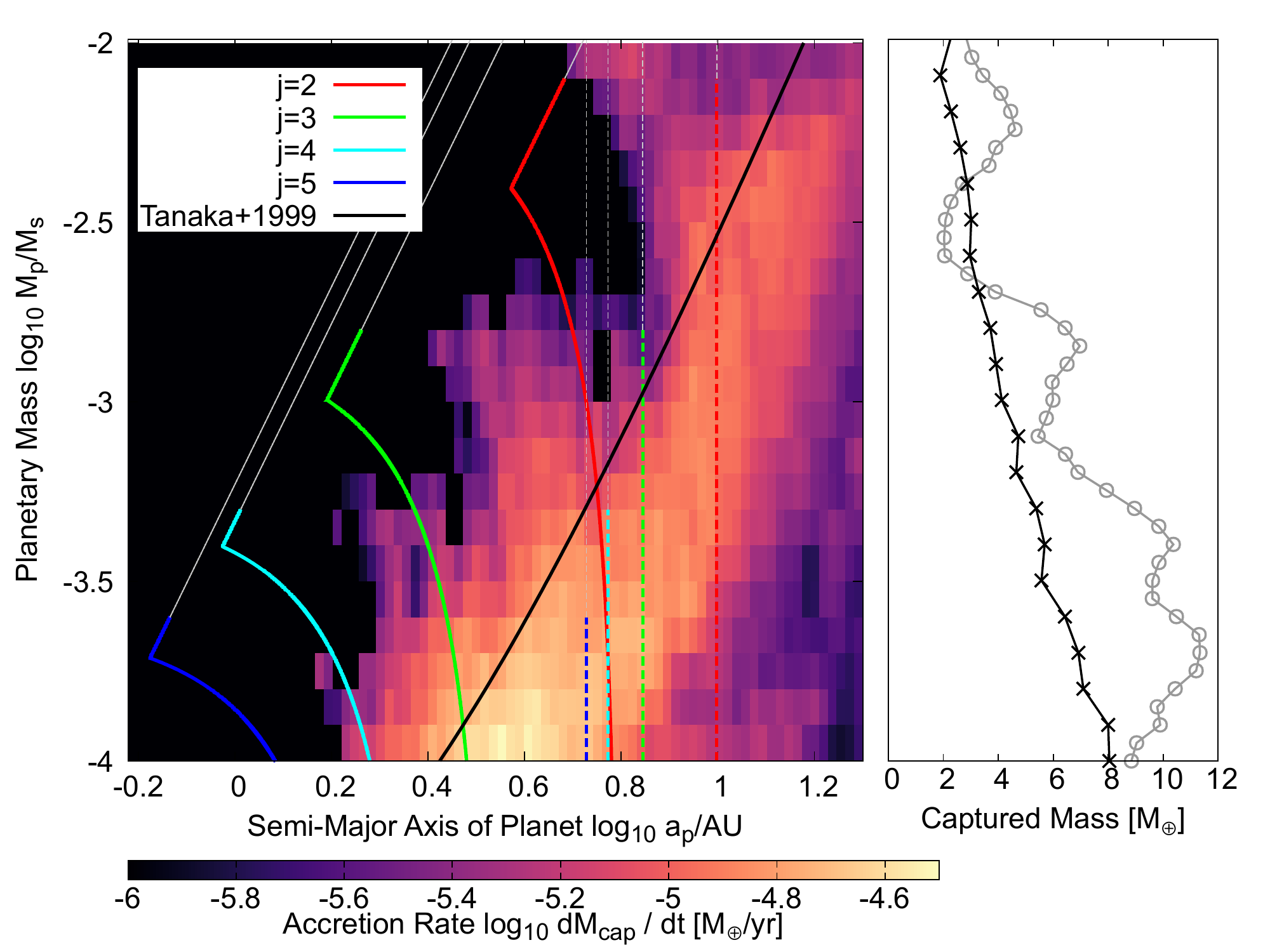}
    \caption{
    Same as Fig.~\ref{fig:Result_ParameterStudy_Rpl}, but for the results of a parameter study for planetary mass $M_{\rm p}$ with the artificial perturbations (see Eq.~(\ref{eq:artificial_perturbation})).
    In the right panel, the results of a parameter study for planetary mass $M_{\rm p}$ without the artificial perturbations are plotted with gray circles for comparison.
    }
    \label{fig:Result_Discussion_Mp_wfp}
  \end{center}
\end{figure}
%%%%%%%%%%%%%%%%%%%%%%%%%%%%%%%%%%%%%%%%%%%%%%%%%%%%%%%%%%%%%

High-velocity collisions of planetesimals perturb the orbit of trapped planetesimals and break the resonant trapping.
Using numerical simulations, \cite{Malhotra1993b} found that for a planet of 10 $M_\oplus$, resonant trapping is broken with a velocity kick of $\gtrsim0.003 v_{\rm K}$. 
Therefore, once $M_{\rm shep}$ exceeds the threshold value and the trapped planetesimals start to collide with each other, the resonant trapping can be broken by planetesimal collisions.
However, the resonant width increases with increasing planetary mass (see Eq.~(\ref{eq:1st_order_resonant_width}); namely breaking the resonant trapping is more difficult for more massive planets) and not all high-velocity collisions lead to the breaking of resonant trapping (see Fig.~6b of \citet{Malhotra1993b}).
In addition, as mentioned in the previous section, the alignment of the orbits by the resonant trapping might reduce the relative velocity of planetesimals.
The possibility of resonant breaking by the planetesimal collisions need to be investigated in future work.
Here, we consider the cases where the resonant trapping is broken by planetesimal collisions.

As shown in sec.~\ref{sec:sweet_spot}, planetesimal accretion occurs only in a limited ring-like region within the protoplanetary disk.
This accretion process seems different from the one found for Earth-mass planets in  \citet{Tanaka+1999}.
In their dynamical simulations, \citet{Tanaka+1999} broke the resonant traps instantaneously by exerting artificial perturbations on the trapped planetesimals (see below), supposing that the planetesimals have their motion changed via the interactions with other planetesimals. Then, comparing the timescale for the protoplanet's migration with that for scattering of planetesimals by the migrating protoplanet, they derived a condition required for planetesimal accretion as: 
\begin{align}
    \tau_{{\rm tide},a} &< 0.81 \left\{ \sqrt{1+0.45 \left(\frac{\tau_{\rm aero,0}}{T_{\rm K,p}} \right)^{2/3} } +1 \right\}^2 \frac{T_{\rm K,p}}{h^{2}}. \label{eq:critical_migration_timescale_Tanaka+1999}
\end{align}
By contrast, when deriving Eq.~(\ref{eq:sweet_spot}) and also performing the dynamical simulations, we considered an extreme case without such modification to the planetesimals motion.

Using the artificial perturbation developed in \citet{Tanaka+1997}, we perform additional numerical simulations.
We randomise the phase angle by adding displacements in orbital elements except $a$, $e$ and $i$. 
The size of the displacement is set as: 
\begin{align}
    \Delta \theta = 2 \pi f_{\rm p}, \label{eq:artificial_perturbation}
\end{align}
where $f_{\rm p}$ is a scaling factor of perturbation strength and randomly given from $-0.01$ to $0.01$.
We add $\Delta \theta$ to orbital angles (longitude of ascending node $\Omega$, longitude of pericentre $\varpi$ and mean longitude at epoch $\epsilon$) every conjunction time.
Figure~\ref{fig:Result_Discussion_time_phase_angle_wfp} shows the temporal change in the phase angle in the case with (a) and without (b) the artificial perturbations.
As shown in panel~(a), even under strong perturbations (corresponding to up to $1~\%$ displacement in the phase angle at every conjunction), the mean motion resonances with a Jupiter-mass planet are efficient in trapping planetesimals and, thereby, resonance shepherding occurs. This is because the resonance width (see Eq.~(\ref{eq:1st_order_resonant_width})) is larger for Jupiter-mass planets than that for Earth-mass planets.
The perturbations is, however, found to hasten the breakup of resonant trapping, because resonant trapping is weakened and over-stable libration is accelerated.

Figure~\ref{fig:Result_Discussion_Mp_wfp} shows the results of the parameter study when varying the planetary mass $M_{\rm p}/M_{\rm s}$ including the artificial perturbations given by Eq.~(\ref{eq:artificial_perturbation}).
For comparison, we also show the results without artificial perturbations with open circles in the right panel. 
Due to the artificial perturbations, the \SSP\ shifts outward (see the left panel) and the total mass of captured planetesimals is reduced (see the right panel). 
In the left panel, the black solid line shows the condition of shepherding obtained in \citet{Tanaka+1999} (see eq.~(\ref{eq:critical_migration_timescale_Tanaka+1999})); planetesimal accretion occurs in regions exterior to it. 
This black line corresponds to the inner boundary of the \SSP\ for the case where resonant traps are perfectly broken.
Although it might seem counter-intuitive, we conclude that without perturbations, resonant trapping shifts the \SSP\ inward and enhances planetesimal accretion due to the effect of accretion bands.

\subsubsection{Ablation of eccentric planetesimals}\label{sec:Discussion_Ablation}
A highly-excited planetesimal has a high relative velocity compared to that of the disk gas.
Strong gas drag from the gaseous disk \citep[e.g.,][]{Pollack+1986} and generated bow-shock \citep[e.g.,][]{Tanaka+2013} heats the planetesimal's surface and triggers the ablation of volatile materials.
Planetesimals trapped in the mean motion resonances have eccentric orbits and feel strong gas drag, and therefore ablation is expected to be important.
The ablation rate depends on the rate of heat transfer from the ambient disk gas $\Lambda$.
We set $\Lambda$ to range between  $0.003$ and $0.6$ for gas drag heating, and between $\sim0.01$ and $\sim0.1$ for bow shock heating \citep[see][and references therein]{Tanaka+2013}.
Recently \citet{Eriksson+2021} explored the possibility of perfect ablation of planetesimals around a  massive protoplanet using $\Lambda=C_{\rm d}/4$, which is the upper limit of the heat transfer rate \citep{DAngelo+2015}.
According to their nominal model result, ablation of highly excited planetesimals is effective in the inner disk region $\lesssim10\AU$ and large amount of solid materials are ablated.
The heating process strongly depends on the eccentricity of planetesimals.
Using eqs.~(\ref{eq:equilibrium_resonant_argument_app}) and (\ref{eq:sweet_spot}), we find that the eccentricity of planetesimals inside the SSP decreases from $\sim0.5$ to $0.1$.
To avoid effective ablation planetesimals must enter the SSP in the outer disk region $\gtrsim10\AU$.
If the secular resonance triggered by the disk's gravity is effective, the ablation would be effective  in more outer disk region \citep{Nagasawa+2019}.
On the other hand, if $\Lambda$ is much smaller than $C_{\rm d}/4$, the ablation of planetesimals would be inefficient.

The investigation presented above concerning planetesimal collision and ablation implies that planetesimal accretion is favoured in the outer disk region $\gtrsim10\AU$. 
Detailed future studies that focus on collision and ablation processes are clearly desirable.

\subsection{The case of Type II migration with shallow-gap}\label{sec:Discussion_Kanagawa}
In the classical picture of the Type II migration \citep[e.g.][]{Lin+1993}, a giant planet migrates with a deep gap opened in the vicinity of its orbit in a gaseous protoplanetary disk. 
For an extremely deep gap, the radial gas flow toward the central star is stemmed completely by the gap and, consequently, the planet migrates with the gaseous disk contracting via viscous diffusion.
Recent hydrodynamic simulations, however, reveal that the gap bottom is not so deep and the gas can cross the gap \citep{Kanagawa+2015,Kanagawa+2016,Kanagawa+2017}.
In this case, the planetary migration is not in the classical regime; instead, the torque exerted on the planet is given similarly to that in the Type I regime \citep{Kanagawa+2018}.
Thus, the migration timescale in the new type II regime is written as: 
\begin{align}\label{eq:type2_migration_regime_K18} 
    \tau_{\rm tide,{\it a},II} &\sim  \frac{1}{2 c} \left(\frac{M_{\rm p}}{M_{\rm s}}\right)^{-1} 
                                \left(\frac{{r_{\rm p}}^2\Sigma_{\rm gap}}{M_{\rm s}}\right)^{-1} 
                                \left(\frac{h_{\rm s}}{r_{\rm p}} \right)^2 {\Omega_{\rm K}}^{-1}, 
\end{align} 
where $c$ is a constant of the order of unity and $\Sigma_{\rm gap}$ is the surface density of disk gas at the gap bottom; $\Sigma_{\rm gap}$ depends on the unperturbed surface density of disk gas $\Sigma_{\rm gas}$ and is given from the hydrodynamic simulations of \citet{Kanagawa+2017} as: 
\begin{align}\label{eq:gap_Kanagawa}
    \Sigma_{\rm gap} = \frac{\Sigma_{\rm gas}}{1+0.04 K},
\end{align}
with
\begin{align}
    K = \left( \frac{M_{\rm p}}{M_{\rm s}} \right)^2 \left( \frac{h_{\rm s}}{r_{\rm p}} \right)^{-5} {\alpha_{\rm vis}}^{-1},
\end{align}
where $\alpha_{\rm vis}$ is the alpha parameter for disk gas turbulent viscosity \citep{Shakura+1973}.
From Eqs.~(\ref{eq:type2_migration_regime_K18}) and (\ref{eq:gap_Kanagawa}), we obtain: 
\begin{align}
    \tau_{{\rm tide},a,{\rm II}} &\sim   \frac{0.02}{c} {\alpha_{\rm vis}}^{-1} \left(\frac{{r_{\rm p}}^2\Sigma_{\rm gas}}{M_{\rm p}}\right)^{-1} 
                                \left(\frac{h_{\rm s}}{r_{\rm p}} \right)^{-3} {\Omega_{\rm K}}^{-1}, \label{eq:type2_migration_regime_simple}
\end{align}
for $1 \ll 0.04 K$ or 
\begin{align}
    M_{\rm p} \gg 8 M_{\oplus} \left( \frac{\alpha_{\rm vis}}{10^{-3}} \right)^{1/2} \left( \frac{r_{\rm p}}{1 \AU} \right)^{5(1-2\beta_{\rm disk})/4}.
\end{align}
Here we discuss the location of SSP and planetesimal accretion onto the protoplanet migrating with eq.~(\ref{eq:type2_migration_regime_simple}).

\subsubsection{The location of the accretion sweet spot}\label{sec:Discussion_SweetSpot}
The location of the \SSP\ depends on the ratio of the planetary migration and aerodynamic gas drag timescales.
The timescale of damping by aerodynamic gas drag $\tau_{\rm aero,0}$ in the disk mid-plane also depends on the disk structure (see Eq.~(\ref{eq:aerodynamic_damping_timescale})) as: 
\begin{align}
    \tau_{\rm aero,0} &= \frac{8\sqrt{2\pi}}{3 C_{\rm d}} \left(\frac{\Sigma_{\rm gas}}{\rho_{\rm pl} R_{\rm pl}}\right)^{-1} \frac{h_{\rm s}}{r_{\rm pl}} {\Omega_{\rm K}}^{-1}. \label{eq:damping_timescale_simple}
\end{align} %
Note that $\tau_{\rm tide,{\it a},II}$ is a function of $r_{\rm p}$, whereas $\tau_{\rm aero,0}$ is a function of $r_{\rm pl}$; $r_{\rm p}/r_{\rm pl}=((j-1)/j)^{2/3}$ for a planetesimal in the $j$:$j-1$ mean motion resonance with the planet.
In this case, Eq.~(\ref{eq:damping_timescale_simple}) can be written as a function of $r_{\rm p}$ as: 
\begin{align}
    \tau_{\rm aero,0} &= \frac{8\sqrt{2\pi}}{3 C_{\rm d}} \left(\frac{\Sigma_{\rm gas}}{\rho_{\rm pl} R_{\rm pl}}\right)^{-1} \frac{h_{\rm s}}{r_{\rm p}} {\Omega_{\rm K}}^{-1} \left(\frac{j-1}{j}\right)^{2\gamma_{\rm disk}/3}, \label{eq:damping_timescale_simple2}
\end{align}
with
\begin{align}
    \gamma_{\rm disk} = 2+\alpha_{\rm disk} -\beta_{\rm disk}. \label{eq:sweet_spot_constant_gamma}
\end{align}
Here, we have assumed that the planetesimals trapped in mean motion resonances are outside the gap.
If the disk viscosity is small, however, the gap slope reaches the locations of the resonances; in that case, the damping timescale increases a few times.
For simplicity, we neglect this effect.

From Eqs.~(\ref{eq:type2_migration_regime_simple}) and (\ref{eq:damping_timescale_simple2}), the ratio of the two timescales is given by: 
\begin{align}
    \frac{\tau_{\rm aero,0}}{\tau_{{\rm tide},a,{\rm II}}} = 
    1 \times10^{-6} C^{-1} \left(\frac{M_{\rm p}}{M_{\rm s}}\right)^{-1} 
                                                                                \left(\frac{r_{\rm p}}{1\AU}\right)^{4(1-\beta_{\rm disk})}
                                                                                \left(\frac{j-1}{j}\right)^{2\gamma_{\rm disk}/3}, \label{eq:fraction_timescales_simple}
\end{align}
with
\begin{align}
    C       &=  \frac{2C_{\rm d}}{c}
                \left(\frac{\alpha_{\rm vis}}{10^{-3}}\right)^{-1}
                \left(\frac{\rho_{\rm pl}}{2 \g/\cm^3}\right)^{-1}
                \left(\frac{R_{\rm pl}}{10^7~\cm}\right)^{-1}
                \left(\frac{M_{\rm s}}{M_{\rm \odot}}\right)
                \left(\frac{\left|h_{\rm s}/r\right|_{\rm 1\AU}}{0.03}\right)^{-4}, \label{eq:sweet_spot_constant_C}
\end{align}
Equation~(\ref{eq:fraction_timescales_simple}) indicates that the surface gas density profile only weakly affects the timescale ratio, which does not explicitly depend on $\Sigma_{\rm gas}$ and weakly depends on $\alpha_{\rm disk}$, namely, $((j-1)/j)^{2\alpha_{\rm disk}/3}$. 
This means that as the protoplanetary disk evolves, the timescale ratio hardly changes, suggesting that the location of the \SSP\ is fixed. % during the planetary formation stage.

%%%%%%%%%%%%%%%%%%%%%%%%%%%%%%%%%%%%%%%%%%%%%%%%%%%%%%%%%%
\begin{figure}%f1
    \begin{center}
    \includegraphics[width=80mm]{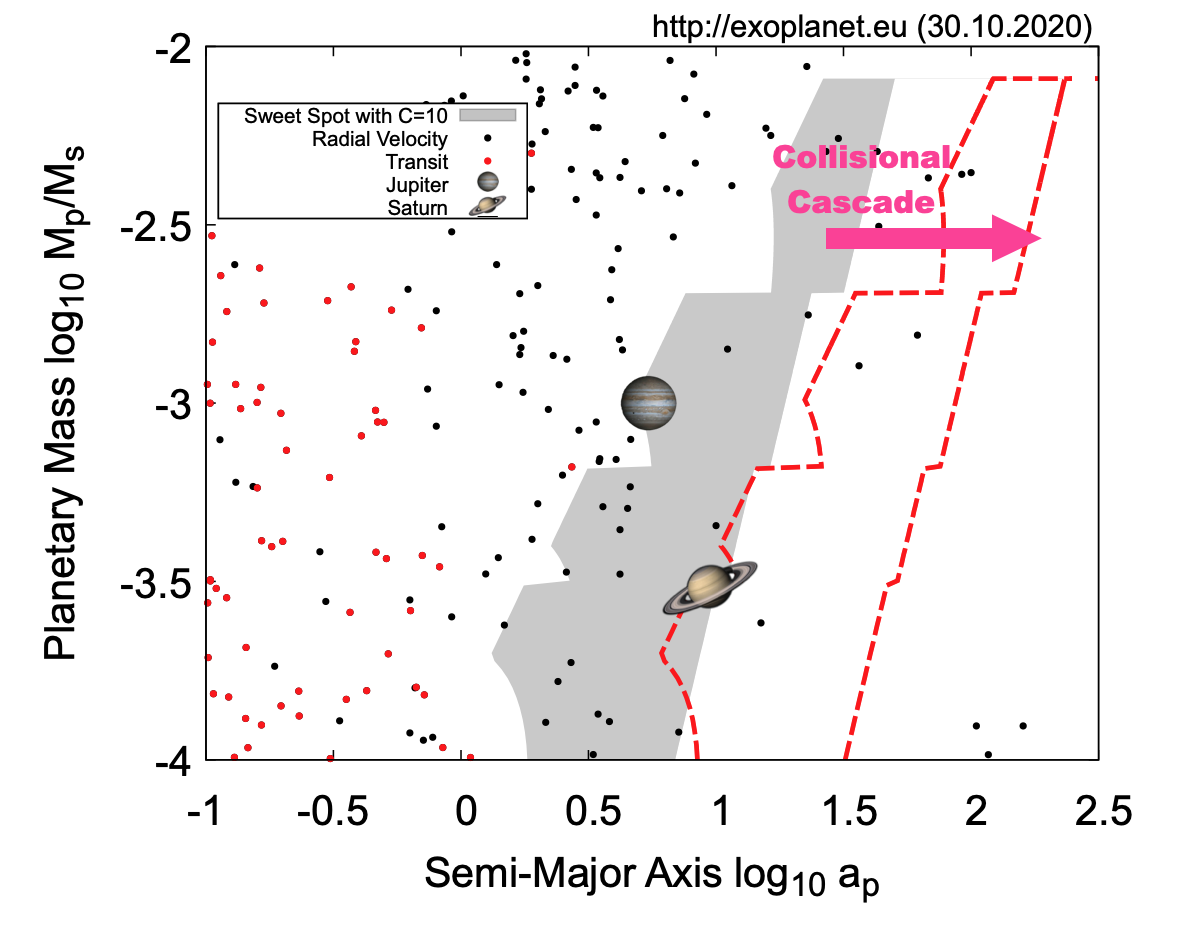}
        \caption{
        Distribution of confirmed exoplanets in the semi-major axis vs. planetary mass plane 
        (http://exoplanet.eu).
        The black points are exoplanets observed by the radial-velocity method.
        The red points are exoplanets observed by the radial-velocity + transit methods.
        As for masses of exoplanets indicated with the black points, the observe values of $M_{\rm p} \sin i$ are simply used.
        The gray area shows the theoretical sweet spot for planetesimal capture given by Eqs.~(\ref{eq:sweet_spot_in_disk_1}) and (\ref{eq:sweet_spot_in_disk_2}) with $C=10$. 
        The red dashed line shows the theoretical sweet spot for planetesimal capture with $C=10^3$, which corresponds to the case where collisional cascade broke planetesimals down to $R_{\rm pl}\lesssim10^5 \cm$. 
        }
        \label{fig:Application_Exoplanet}
    \end{center}
\end{figure}
%%%%%%%%%%%%%%%%%%%%%%%%%%%%%%%%%%%%%%%%%%%%%%%%%%%%%%%%%%

Assuming $\alpha_{\rm disk}$ = 1, which is valid in a steady viscous accretion disk except for its outermost region 
\citep[e.g.][]{Lynden-Bell+1974}, and $\beta_{\rm disk} = 1/4$, which is valid in an optically thin disk, and substituting Eq.~(\ref{eq:fraction_timescales_simple}) into Eq.~(\ref{eq:sweet_spot}), we obtain the semi-major axes of the inner and outer edge of \SSP\ , $a_{\rm SS,in}$ and $a_{\rm SS,out}$, respectively, as: 
\begin{align}
    a_{\rm SS,in} &= {\rm max} \left.
            \begin{cases}
                1 C^{1/3} \left(\frac{M_{\rm p}/M_{\rm s}}{10^{-3}}\right)^{2/3} \left(\frac{j}{j-1} \right)^{10/9} \\%1.4 
                1 C^{1/3} \left(\frac{M_{\rm p}/M_{\rm s}}{10^{-3}}\right)^{1/3} \left\{\frac{j^{11}}{(j-1)^5} \right\}^{1/18} \frac{e_{\rm cross}}{0.1}
            \end{cases} \right\} \AU, \label{eq:sweet_spot_in_disk_1} \\%1.1 
    a_{\rm SS,out} &= 8 C^{1/3} \left(\frac{M_{\rm p}/M_{\rm s}}{10^{-3}}\right)^{1/3} \left\{ \frac{j^{11}}{(j-1)^{14}} \right\}^{1/18} \AU. \label{eq:sweet_spot_in_disk_2}%7.6
\end{align}

Figure~\ref{fig:Application_Exoplanet} shows exoplanets identified so far and the predicted location of the \SSP\ in the $a_{\rm p}$-$M_{\rm p}$ plane for {\bf $C=10$}.
The \SSP\ locates from $\sim$ 1~$\AU$ to $\sim$ 30~$\AU$ and shifts outward with increasing planetary mass.
The red and black symbols show exoplanets observed using the transit method and the radial velocity method, respectively (http://exoplanet.eu).
Given that planetary growth occurs via the runaway gas accretion, followed by inward migration, the evolution path of a gas giant planet can be drawn from right-bottom to left-top direction in the $a_{\rm p}$-$M_{\rm p}$ plane \citep[e.g.,][]{Ida+2004a,Mordasini+2009a}. 
If runaway gas accretion begins in outer disk region, which might be favoured for planetary core formation because of the increasing isolation mass of protoplanet \citep[e.g.][]{Armitage+2010}, the exoplanets currently observed in the region interior to the \SSP\ must have crossed the \SSP\ during their formation stages. 

\subsubsection{Maximum mass of planetesimals shepherded into the SSP}\label{sec:Discussion_SweetSpot_Mshep}
%%%%%%%%%%%%%%%%%%%%%%%%%%%%%%%%%%%%%%%%%%%%%%%%%%%%%%%%%%%%%%%%
\begin{figure}%f1
  \begin{center}
    \includegraphics[width=80mm]{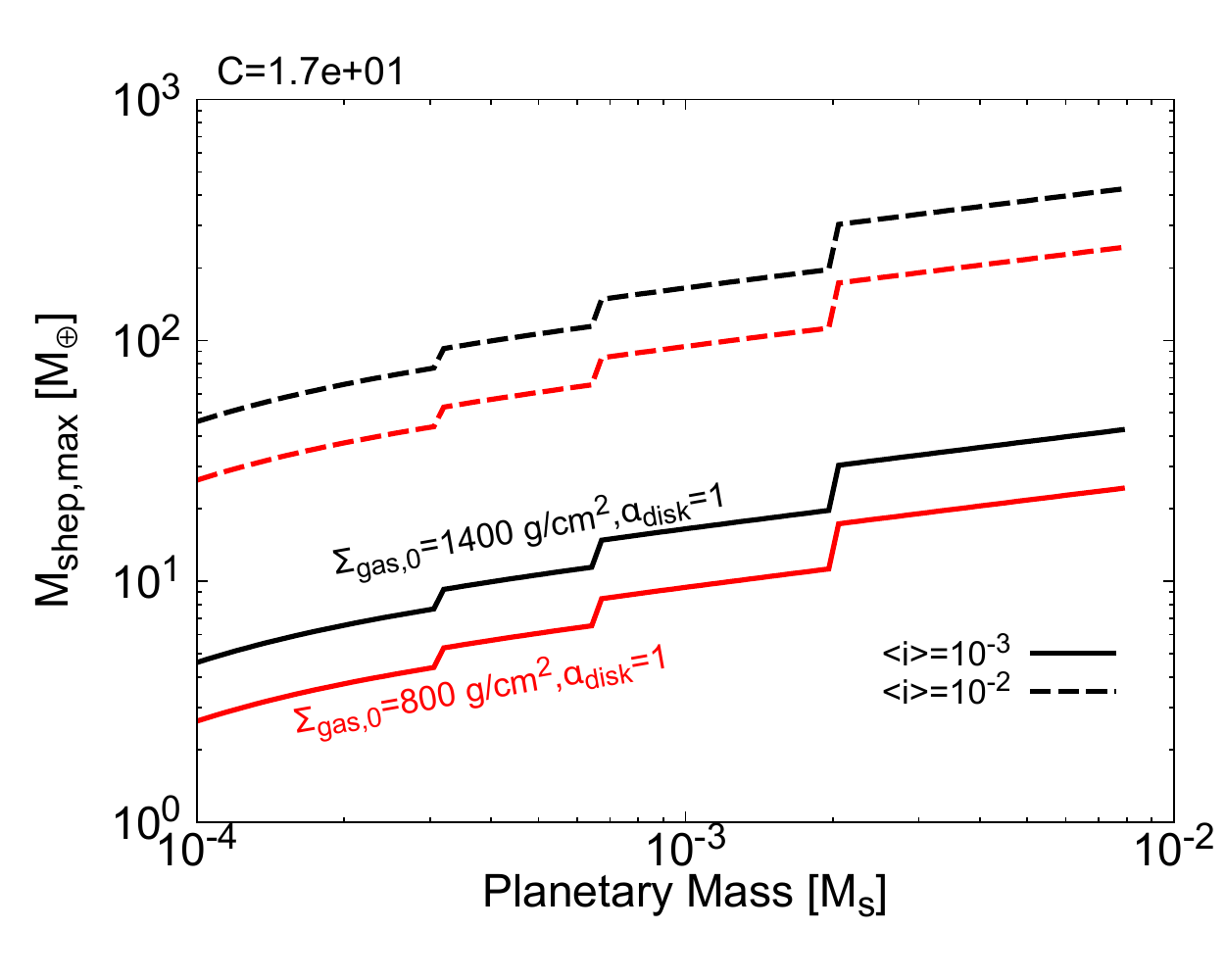}
    \caption{
    The maximum mass of planetesimals shepherded into the SSP without active planetesimal collisions $M_{\rm max,shep}$ as a function of planetary mass $M_{\rm p}$.
    Here, we consider the case where the protoplanet migrates with the type II regime obtained by \citet{Kanagawa+2018} (see eq.~(\ref{eq:type2_migration_regime_K18})).
    The solid and dashed lines show the cases of $\left< i \right>=10^{-3}$ and $10^{-2}$, respectively.
    The black and red lines are cases of massive disk with $\Sigma_{\rm gas,0}=1400 \g/\cm^2$ and $800 \g/\cm^2$, respectively.
    We adapt the optically thick disk of $T_{\rm 0}=140 \K$, the disk viscosity $\alpha_{\rm vis}=10^{-3}$, the size of planetesimal $R_{\rm pl}=10^7 \cm$ and the planetesimal density $\rho_{\rm pl}=1 \g/\cm^3$.
    The scaling parameter of the location of the SSP $C$ is obtained as $C=17$.
    }
    \label{fig:Mp_Mshep}
  \end{center}
\end{figure}
%%%%%%%%%%%%%%%%%%%%%%%%%%%%%%%%%%%%%%%%%%%%%%%%%%%%%%%%%%%%%%%%
The total mass of accreted planetesimals depends on the amount of planetesimals that  shepherded into the SSP. 
By substituting eqs.~(\ref{eq:resonant_center}), (\ref{eq:type2_migration_regime_simple}) and (\ref{eq:sweet_spot_in_disk_2}) into eq.~(\ref{eq:maximum_Mshep}), we can estimate $M_{\rm shep,max}$ when the protoplanet reaches the outer edge of the SSP.
In this case, $M_{\rm shep,max}$ weakly depends on $\alpha_{\rm vis}$, $\rho_{\rm pl}$, $R_{\rm pl}$ and ${M_{\rm p}}^{-1}$ ($-1/4$ powers of each value), but linearly depends on $\tan \left< i \right>$, $\Sigma_{\rm gas,0}$ and ${T_{\rm 0}}^{-1}$.
Here, we adapt the optically thick disk of $T_{\rm 0}=140\K$ \citep{Sasselov+2000} and consider two disks with  $\Sigma_{\rm gas,0}=1400 \g/\cm^2$ and $800 \g/\cm^2$.
The former disk corresponds to the massive disk of total mass $0.1 M_{\rm s}$ and typical radius $100\AU$.
Figure~\ref{fig:Mp_Mshep} shows $M_{\rm shep,max}$ as a function of planetary mass $M_{\rm p}/M_{\rm s}$.
We find that a few tens $M_{\oplus}$ of planetesimals can be shepherded into the SSP before initiating collisional cascade even with $\tan\left< i \right>=10^{-3}$.
For the capture of several tens or more planetesimals,  the effect enhancing $\tan\left< i \right>$ needs to be considered. 

Note that $M_{\rm shep,max}$ only provides the maximum mass of planetesimals shepherded into the SSP and not the mass of  planetesimals that can be captured by the protoplanet.
The numerical simulations we presented in sec.~\ref{sec:sweet_spot_numerical} suggest that $\sim 30\%$ of the shepherded planetesimals are captured by the protoplanet. 
This accretion efficiency, however, depends on various parameters such as the inclination of planetesimals \citep{Tanaka+1999} and the planetary capture radius \citep{Inaba+2003}.
In addition, the initial semi-major axis of planetesimals also affects the capture efficiency because of the mean motion resonances. 
A detailed investigation of the accretion efficiency under various conditions should be conducted in future research.

\subsection{Heavy-element enrichment of giant planets}\label{sec:Application_Model_InnerPlanet_result}
If giant planets cross the sweet spot during the migration phase, planetesimal accretion could take place.
If tens Earth-masses of planetesimals were shepherded into the SSP as shown in fig.~\ref{fig:Mp_Mshep}, we would expect a clear difference in the total heavy-element mass of planets depending on whether they migrated interior or exterior to the \SSP.
If such a difference between planetary populations is observed, it could confirm that planetesimal accretion is a significant process that takes place during planetary migration. 
However, once the collisional cascade occurs, the location of the SSP moves to the outer disk region and the metallicity difference between the populations would be reduced.
It is therefore still unclear whether such a difference is detectable. 

Currently, there are no estimated  metallicities for giant exoplanets within the outer disk region (i.e., at distances  $\gtrsim 10\AU$).
To estimate the metallicity of giant exoplanets, measurements of the radius, mass and age of the planet are required. 
At the moment, data of the masses and radii of cold giant planets are still unavailable.
Nevertheless, the upcoming decades are expected to provide vast observations thanks to various missions.
Wide field observations, such as TESS and Plato, will increase the number of cold giant planets \citep{Ricker+2014,Rauer+2014}.
Follow-up observations using precise radial velocity measurement, such as HARPS \citep{Mayor+2003} and ESPRESSO \citep{Pepe+2010}, and precise transit measurement, such as NGTS \citep{Wheatley+2018} and WASP \citep{Pollacco+2006}, are suitable for obtaining the masses and radii of giant planets at large radial distances. 
In addition, high-dispersion coronagraphy technique \citep[e.g.][]{Kotani+2020} could  constrain the atmospheric composition of cold giant planets.
Next generation observations will constrain the main source of heavy elements.

In this study, we focus on planetesimal accretion and consider it as the main source of heavy-element enrichment in gas giant planets.
However, other various processes could also lead to heavy-element enrichment. 
The metallicity of the gaseous disk can be enriched by the sublimation of dust grains and pebbles crossing each ice lines \citep{Booth+2017,Booth+2019}. 
The accretion of these enriched disk gas brings massive heavy elements into gas giant planets \citep{Schneider+2021}.
Giant impacts have also been investigated as a source of enrichment  \citep{Ikoma+2006b,Liu+2015,Liu+2019,Ginzburg+2020, Ogihara+2021}.
By combining the effects of pebble accretion and giant impacts, \citet{Ogihara+2021} found that up to $\sim100~M_{\oplus}$ of heavy elements can be accreted by a giant planet through multi-giant impacts of Neptune-mass planets.

In order to determine what is the main source of the heavy-element enrichment in giant exoplanets, other observational information such as elemental ratios (refractory-to-volatile, C/O, O/H, etc) is valuable. 
In recent observations, the existences of TiO and VO in Hot-Jupiters is suggested by the transmission spectrum analysis \citep{Evans+2016,Chen+2021}.
The amount of these refractory materials is useful for constraining the dominant process of heavy-element accretion.
The accretion of massive refractory materials is only possible through solid accretion, thus using the information of refractory materials, we can constrain whether the heavy elements are brought by solid accretion or gas accretion.

The elemental ratio of volatile materials also provides information on the accretion process of heavy-elements \citep[e.g.,][]{Madhusudhan+2014,Notsu+2020,Turrini+2021,Schneider+2021}.
\citet{Turrini+2021} investigated the effect of planetesimal accretion on elemental ratios of volatile materials such as C/O, N/O or S/N.
They considered a Jupiter-mass planet in a typical disk model, however, the relative position between the \SSP\ and each ice lines changes with the planetary mass and disk conditions.
The elemental ratio of planetesimals shepherded into the  \SSP\ would depend on the evolution pathways of the giant planets on the $a_{\rm p}$-$M_{\rm p}$ plane and disk conditions.
We hope to further investigate the link between elemental ratios and the formation history in followup studies. 
This is highly relevant since future missions like JWST and Ariel will provide accurate measurements of the atmospheric composition of gaseous planets that can then be used to better understand the planetary origin.

\section{Summary \& Conclusions}
\label{sec:conclusion}
Planetesimal accretion during planetary migration could lead to a significant enrichment of giant planets with heavy elements. 
As revealed in \citet{Shibata+2020}, planetesimal accretion occurs in a sweet spot for planetesimal accretion (\SSP), where planetesimals can be captured by the migrating planet efficiently.
By performing dynamical simulations for planetesimals with a migrating planet, \citet{Shibata+2020} demonstrated that the \SSP\ emerges due to shepherding processes caused by aerodynamic gas drag and mean motion resonances.
However, they conducted no detailed analysis of the nature (especially, the location) of the \SSP\ and therefore in this paper we have focused on the nature of the accretion sweet spot. 

In sec.~\ref{sec:sweet_spot}, we analytically derived an equation describing the location of the \SSP and found the following: 
(i) The  \SSP\ is regulated by the ratio of the damping timescale for the planetesimal eccentricity due to aerodynamic gas drag to the migration timescale of the planet  
(ii) The \SSP\ also depends on the planetary mass, because the accretion bands change with the relative positions between the mean motion resonances and the feeding zone.

In sec.~\ref{sec:sweet_spot_numerical}, we performed numerical simulations and confirmed that the derived conditions reproduce the numerical results well.
In sec.~\ref{sec:discussion}, we discussed the effect of planetesimal collisions and ablation during the shepherding process.
These effects would be important for inner disk region as $\lesssim10\AU$ and put other constraint for planetesimal accretion process.
%{\bf
%We suggest that the mutual collisions of planetesimals in the resonant shepherding is important for determining the population of small objects in planetary systems. 
%}
If the migration timescale is inversely proportional to the disk gas surface density 
\citep[e.g.,][]{Kanagawa+2018}, the timescale ratio is rather insensitive to the structure of the protoplanetary disk.
Therefore, the location of the \SSP\ is fixed during the evolution of the protoplanetary disk.

Finally, we discussed the effect of the \SSP\ on the heavy-element enrichment of gas giant planets.
We suggest that the existence of the \SSP\ makes a difference in the heavy-element mass between planets observed in the regions interior and exterior to the \SSP.
The detailed chemical composition of gas giant planets, such as refractory-to-volatile ratio or C/O ratio, would be also affected by the location of the \SSP.
If such compositional features can be observed, it could be a piece of the evidence of planetesimal accretion during planetary migration.
Future observations in the upcoming decades will reveal whether planetesimal accretion is a main source of heavy elements in giant planets, and therefore advance our understanding of their origin.

\begin{acknowledgements}
      S.~S.~ and R.~H.~acknowledge support from the Swiss National Science Foundation (SNSF) under grant \texttt{\detokenize{200020_188460}}. 
      Part of this work was supported by the German
      \emph{Deut\-sche For\-schungs\-ge\-mein\-schaft, DFG\/} project
      number Ts~17/2--1 and by JSPS Core-to-Core Program “International Network of Planetary Sciences (Planet$^2$)” and JSPS KAKENHI Grant Numbers 17H01153 and 18H05439. 
\end{acknowledgements}

\appendix
\section{Model descriptions}
\label{app:method}
\subsection{Equations of motion}\label{sec:app_eom}
The equation of motion is given by: 
\begin{align}
   \frac{{\rm d} {\bf r}_i }{{\rm d} t} &= \sum_{i\neq j} {\bf f}_{{\rm grav},i,j} + {\bf f}_{\rm aero} +{\bf f}_{\rm tide}, \label{eq:EoM}
\end{align}
where ${\bf r}_i$ is the position vector relative to the initial (i.e., $t = 0$) mass centre of the star-planet-planetesimals system, %$t$ is the time, 
${\bf f}_{{\rm grav},i,j}$ is the mutual gravity between particles $i$ and $j$ given by: 
\begin{align}
   {\bf f}_{{\rm grav},i,j} &= - \mathcal{G} \frac{M_j}{{r_{i,j}}^3} {\bf r}_{i,j} \label{eq:EoM_gravity}
\end{align}
with $\mathbf{r}_{i,j}$ being the position vector of particle $i$ relative to particle $j$ ($r_{i,j} \equiv |{\bf r}_{i,j}|$),
$M_{j}$ is the mass of particle $j$, and $\mathcal{G}$ is the gravitational constant. 
${\bf f}_{\rm aero}$ is the aerodynamic gas drag, and ${\bf f}_{\rm tide}$ is the gravitational tidal drag from the protoplanetary disk gas. 
The central star, planet, and planetesimals  are denoted by the subscripts $i$ (or $j$) = 1, 2, and $\geq$~3, respectively.
The planetesimals are treated as test particles; therefore $f_{{\rm grav},i,j}=0$ in Eq.~(\ref{eq:EoM}) for $j \geq~3$.
The aerodynamic gas drag force and the tidal drag force are given respectively by:  
\begin{align}
  {\bf f}_{\rm aero} &= -\frac{{\bf u}}{\tau_{\rm aero}}, \label{eq:EoM_reduced_gas_drag} \\
  {\bf f}_{\rm tide} &= - \frac{{\bf v}_{\rm p}}{2 \tau_{{\rm tide},a}},
\end{align}
with
\begin{align}
    \tau_{\rm aero} &=  \frac{2 m_{\rm pl}}{ C_{\rm d} \pi R_{\rm pl}^2 \rho_{\rm gas} u} = \tau_{\rm aero, 0} \frac{v_{\rm K}}{u}% \\
%    \tau_{{\rm tide},a} &= \tau_{\rm tide,0} \left( \frac{a_{\rm p}}{1 \AU} \right)^{1/2}. \label{eq:type2_migration_timescale}
\end{align}
Here ${\bf u}={\bf v}_{\rm pl}-{\bf v}_{\rm gas}$ ($u = |{\bf u}|$) is the planetesimal's  velocity (${\bf v}_{\rm pl}$) relative to the ambient gas (${\bf v}_{\rm gas}$), %$\tau_{\rm aero}$ is the timescale of aerodynamic gas drag, 
%$m_{\rm pl}$ is the planetesimal's mass, 
%$C_{\rm d}$ is the non-dimensional drag coefficient and given as $C_{\rm d}=1$, 
%$\rho_{\rm gas}$ is the gas density,
%$R_{\rm pl}$ is the planetesimal's radius,
%${\bf v}_{\rm p}$ is the planet's velocity,
%$a_{\rm p}$ is the planet's semi-major axis,
%and $\tau_{\rm tide,0}$ is a scaling parameter.
%The velocity and density of the ambient disk gas are calculated from the protoplanetary disk model.
Given the range of the planetesimal mass ($\sim10^{16}$-$10^{22} \g$) and planet ($\sim10^{30} \g$), we assume $f_{\rm tide}=0$ for the former and $f_{\rm aero}=0$ for the latter.
The central star is not affected by $f_{\rm aero}$ and $f_{\rm tide}$.

\subsection{Gas disk}\label{sec:app_disk}
\begin{comment}
To focus on the dynamic process, we adopt simply the minimum-mass solar nebula model \citep{Hayashi1981} as our gas disk model. 
We assume that the disk has an axisymmetric, temporally unchanged structure and disregard the difference between the locations between the symmetry axis and the mass centre of the star-planet system.
We also neglect the disk's self-gravity  %meaning no gravitational interaction but 
but include the tidal drag  between the disk and planet.
The surface density $\Sigma_{\rm gas}$ is given by: 
\begin{align}\label{eq:Disk_Gas_MMSN}
    \Sigma_{\rm gas} = \Sigma_{\rm gas,0} \left( \frac{r}{1\AU} \right)^{-\alpha_{\rm disk}},
\end{align}
where $r$ is the radial distance from the initial mass centre of the star-planet system, $\Sigma_{\rm gas,0} = 1.7 \times 10^3 \g/\cm^2$ and $\alpha_{\rm disk}=3/2$.
The disk' temperature $T_{\rm disk}$ is given by: 
\begin{align}
    T_{\rm disk} = T_{\rm disk,0} \left(\frac{r}{1\AU} \right)^{-2\beta_{\rm disk}},
\end{align}
where $T_{\rm disk,0}=280~\K$ and $\beta_{\rm disk}=1/4$.
\end{comment}
The protoplanetary disk is assumed to be vertically isothermal, and the gas density $\rho_{\rm gas}$ is expressed as: 
\begin{align}\label{eq:Disk_Gas_volume}
   \rho_{\rm gas} = \frac{\Sigma_{\rm gas}}{\sqrt{2 \pi} h_{\rm s}} \exp \left( -\frac{z^2}{2 {h_{\rm s}}^2} \right),
\end{align}
where $z$ is the height from the disk mid-plane and $h_{\rm s}$ is the disk's scale height.
The aspect ratio of the protoplanetary disk is given by:  
\begin{align}
    \frac{h_{\rm s}}{r} = \frac{c_{\rm s}}{r\Omega_{\rm K}},
\end{align}
where $\Omega_{\rm K}$ is the Kepler angular velocity and $c_{\rm s}$ is the isothermal sound speed of disk gas.
The sound speed is given as %Assuming locally isothermal case, 
\begin{align}
    c_{\rm s} = \sqrt{\frac{k_{\rm B} T_{\rm disk}}{\mu_{\rm disk} m_{\rm H}}} \propto r^{-\beta{\rm disk}},
\end{align}
where $k_{\rm B}$ is the Boltzmann constant and $\mu_{\rm disk}$ is the mean molecular weight in the unit of proton mass $m_{\rm H}$.
In this study, $\mu_{\rm disk}$ is set as $2.3$ and the aspect ratio of the disk gas at $1\AU$ is $\sim0.03$.

The gas in the protoplanetary disk rotates with a sub-Keplerian velocity because of pressure gradient; namely, $v_{\rm gas} = (1-\eta_{\rm gas}) r \Omega_{\rm K}$, where $\eta_{\rm gas}$ is $\ll 1$ and given as
\begin{align}
	\eta_{\rm gas} &\equiv	\mathbf{-} \frac{1}{2} \left( \frac{h_{\rm s}}{r} \right)^2 \frac{{\rm d} \ln P_{\rm gas}}{{\rm d} \ln r}, \label{eq:eta_gas_definition} \\ 
	               &=		\frac{1}{2} \left( \frac{h_{\rm s}}{r} \right)^2 
	                        \left[ \frac{3}{2} \left( 1 - \frac{z^2}{{h_{\rm s}}^2} \right) +\alpha_{\rm disk} +\beta_{\rm disk} \left(1+\frac{z^2}{{h_{\rm s}}^2}  \right) \right];
\end{align}
$P_{\rm gas}$ is the gas pressure.
%For deriving the above equation, we assume $\eta_{\rm gas} \ll 1$ and use the ideal-gas relation for isothermal sound speed, i.e., $c_{\rm s}^2$ = $P_{\rm gas} / \rho_{\rm gas}$.

\subsection{Planetesimals}
\begin{comment}
For the planetesimals we adopt a simple surface density profile given by: %$\Sigma_{\rm solid}$, which is expressed as
\begin{align}
    \Sigma_{\rm solid} = \Sigma_{\rm solid,0} \left( \frac{r}{1\AU} \right)^{-\alpha_{\rm disk}^{\prime}}, \label{eq:Model_Disk_Solid_Surface_Density}
\end{align}
and set $\alpha_{\rm disk}^{\prime}=\alpha_{\rm disk}=3/2$ and 
\begin{align}
    \Sigma_{\rm solid,0} = Z_{\rm s} \Sigma_{\rm gas,0},    
\end{align}
where $Z_{\rm s}$ is the solid-to-gas ratio (or the metallicity).
We assume that $Z_{\rm s}$ equals the metallicity of the central star.
The planetesimal's mass $m_{\rm pl}$ is calculated as $4\pi\rho_{\rm pl}{R_{\rm pl}}^3/3$, where $\rho_{\rm pl}$ is the planetesimal's mean density and given as $\rho_{\rm pl}=2 \g/\cm^3$.
\end{comment}
The surface number density of planetesimals is given as $\Sigma_{\rm solid}/m_{\rm pl}$ which gives $\sim10^6~/\AU^2$ at maximum.
To speed up the numerical integration, we follow the orbital motion of super-particles.
%The super-particles are distributed in a given radial region, the inner and outer radii of which are denoted by $a_{\rm pl,in}$ and $a_{\rm pl,out}$, respectively. 
The surface number density of super-particles $n_{\rm s}$ is given as: 
\begin{align}
    n_{\rm s} = n_{\rm s,0} \left( \frac{r}{1 \AU} \right)^{-\alpha_{\rm sp}}
\end{align}
where
\begin{align}
    n_{\rm s,0} = \frac{N_{\rm sp}}{2 \pi} \frac{2 -\alpha_{\rm sp}}
                  {\left(a_{\rm pl,out}/1\AU\right)^{2-\alpha_{\rm sp}} -\left(a_{\rm pl,in}/1\AU\right)^{2-\alpha_{\rm sp}} } \left[\frac{1}{\AU^2}\right];
\end{align}
$N_{\rm sp}$ is the total number of super-particles used in a given simulation.
In our simulation, we set $N_{\rm sp}=12,000$ and $\alpha_{\rm sp}=1$ to distribute super-particles uniformly in the radial direction.
The mass per super-particle $M_{\rm sp}$ is given by: 
\begin{align}
    M_{\rm sp} (a_{\rm 0}) = \frac{\Sigma_{\rm solid} (a_{\rm 0})}{n_{\rm s} (a_{\rm 0})} ,
\end{align}
where $a_{\rm 0}$ is the initial semi-major axis of the super-particle.

As for eccentricity and inclination, although planetesimals are treated as test particles in our simulations, assuming that planetesimals are, in reality, scattered by their mutual gravitational interaction, we adopt the Rayleigh distribution as the initial eccentricities $e$ and inclinations $i$ of planetesimals.
We set $\left< e^2 \right>^{1/2}=2\left< i^2 \right>^{1/2}=10^{-3}$.
The orbital angles $\Omega$, $\varpi$ and $\epsilon$ are distributed uniformly.

During the orbital integration, we judge that a super-particle has been captured by the planet once (i) the super-particle enters the planet's envelope or (ii) its Jacobi energy becomes negative in the Hill sphere.
The planet's radius $R_{\rm p}$ is calculated as: 
\begin{align}\label{eq:Rplanet_cap}
    R_{\rm p} = \left( \frac{3 M_{\rm p}}{4 \pi \rho_{\rm p}} \right)^{1/3},
\end{align}
where $\rho_{\rm p}$ is the planet's mean density.
The planet's radius is assumed to be  extended shortly after formation and the density is set to  $\rho_{\rm p}=0.125~\g~\cm^{-3}$, corresponding to a planet's radius twice as large as Jupiter's current radius.


%-------------------
\bibliographystyle{aa} % 
\bibliography{refs} %
%-------------------

\end{document}